\documentclass[aip,pop,reprint,groupaddress,noshowpacs,noshowkeys]{revtex4-2}

\usepackage{amsmath,amssymb,mathrsfs}

\usepackage{graphicx}
\usepackage{dcolumn}
\usepackage{bm}
\usepackage{tabularray}
\usepackage{xcolor}
\usepackage[pdftex,
  hidelinks,
  colorlinks=true,
  linkcolor=blue,
  anchorcolor=black,
  citecolor=blue,
  urlcolor=blue,
  pdfauthor= {Michael Glinsky},
  pdfsubject = {Heisenberg Scattering Transformation},
  pdfdisplaydoctitle,
  bookmarks=true,
  bookmarksopen=true]{hyperref}

\begin{document}
\title{A transformational approach to collective behavior}

\author{Michael E. Glinsky}
\affiliation{BNZ Energy Inc., Santa Fe, NM, USA}

\begin{abstract}
This paper presents a new approach to the characterization, forecast, and control of collective systems. Collective systems are an ensemble of conservatively interacting entities. The evolution of the entities are determined by symmetries of the entities. Collective systems take many different forms. A plasma is a collective of charged particles, a fluid is a collective of molecules, a elementary field is a collective of elementary particles, and a cosmos is a collective of celestial bodies. Our new theory builds on the canonical transformation approach to dynamics. This approach recognizes that the symmetry leads to the conservation of a real function, that is the infinitesimal generator of a Lie group. The finite generator of the canonical transformation is derived from the infinitesimal generator by the solution of the Hamilton-Jacobi equation. This generating function is also known as the action, the entropy, and the logarithmic likelihood. The new theory generalizes this generating function to the generating functional of the collective field. Finally, this paper derives the formula for the Mayer Cluster Expansion, or the S-matrix expansion of the generating functional. We call it the Heisenberg Scattering Transformation (HST). Practically, this is a localized Fourier Transformation, whose principal components give the singularity spectrums, that is the solution to the Renormalization Group Equations. Limitations on the measurement of the system (that is the Born Rule and the Heisenberg Uncertainty Principle) lead to quantization of the stochastic probabilities of the collective field. How different collective systems couple together to form systems-of-systems is formalized. The details of a practical implementation of the HST will be presented.
\end{abstract}

\maketitle

\section{Introduction}
\label{intro.sec}
The best introduction to this new theory is a 1982 interview of Paul Dirac by Friedrich Hund.\citep{dirac.82}   Dirac argued that symmetry is primal, and that he preferred the perspective of Einstein's gravitation being a ``a question of geometry'' (i.e., symmetry and topology with systems following geodesic motion).  Although, Dirac was central to the development of Field Theory, he did not like how it was a ``question of probability'', so that the ElectroMagnetic (EM), weak and strong fields could not be unified with gravity.  Einstein shared Dirac's criticism of Field Theory, famously saying, ``God does not play dice''.

The path that we follow is that of Gell-Mann,\citep{gell.mann.00} saying the three forces are a question of group symmetries (that is, the complex Lie groups SU(1), SU(2) and SU(3)).  Dirac in his later career did not realize that he found the key to the lock when he wrote down the infinitesimal group generator for gravity \citep{dirac.58,dirac.59} (that is, the Hamiltonian), $H_g$, leading to the complex Lie group $\mathbb{H}_g = \mathscr{H}_g \otimes \text{Ad}(\mathscr{H}_g)$.  

Dirac also did not like Wilson renormalization,\citep{wilson.71} which he called ``not a logical mathematical process''.  He characterized it as ``a set of working rules, rather than a correct mathematical theory''.  Wilson himself recognized the deficiencies in his theory.  He could not find \emph{``a quantitative characterization of a complete orthonormal set of minimal wave packets.  For quantitative purposes one would have to take into account tails of the wave packets which extend outside their assigned cells.  It will be assumed here that one can divide phase space into cells of unit volume in any way one pleases and still be able to construct a corresponding set of minimal wave packets.  There is no guarantee that this is actually possible, and no examples of such a set of wave packets will be given here.''} 

There have been a set of mathematical analysis methods developed, starting in the 1990's with wavelet analysis,\citep{mallat99} that have been laying the path to the solution that we developed which is ``a mathematically logical process'' of renormalization, with a physical interpretation.  Wavelet analysis is based on convolution of a signal with a bank of scaled wavelets.  The relation of the wavelet analysis to the ``singularity spectrums'' of the system was recognized by \citet{herrmann.97} and several other authors.  Wavelet decompositions and their moments were used as parameters of probability distributions that were estimated as Gaussian distributions,\citep{strauss.03,glinsky.et.al.03} and were estimated by Probabilistic Neural Networks (PNNs) as sums of Gaussian distributions.\citep{glinsky.01}  Wavelet analysis and its moments was extended and formalized to an iterative deep convolution by the Mallat Scattering Transformation (MST).\citep{mallat.12}  Between each convolution is the nonlinear activation function, modulus.  It was noted by Mallat and coworkers that the analysis became near linear if the logarithm was taken before and after the transformation.\citep{Bruna2013}  There have been further developments by Mallat and coworkers to include phase, in a rather Byzantine way, with Wavelet Phase Harmonics (WPH);\citep{mallat2020phase,zhang2021maximum,allys2020new,regaldo23}  and to relate it to Wilson renormalization with Wavelet Conditional Renormalization Group (WC-RG).\citep{marchand22}  

The path to the solution has also been laid by the success of Generative Artificial Intelligence (genAI).  This analysis is a deep convolution like the MST, but unlike Mallat, a large optimization over a trillion parameter Banach space is done, followed by a decoding by an autoencoding Multi-Layer Perceptron (MLP).  Two major variants of genAI, Deep Reinforcement Learning (DRL) of David Silver at DeepMind,\citep{mnih15,goodfellow16,sutton18,yoon21,wang21} and Generative Pretrained Transformers (GPTs) of Alec Radford at openAI,\citep{radford.18,farimani23} have the structure of a deep convolution followed by an autoencoder to a Reduced Order Model (ROM) leading to a ``generating'' function.  The generating function in the case of DRL is the approximate value function, $\widetilde{V}_\theta(s)$, the solution to the Hamilton-Jacobi equation.  In the case of GPTs, the generating function is the approximate logarithmic likelihood, $\ln(\widetilde{\rho}_\theta(x))$.  

\citet{glinsky23} started to put the picture together when they assembled a workflow using the MST and WPH as the generating functional or deep deconvolution, and a Multi-Layer Perceptron (MLP) \citep{haykin.98} with Rectified Linear Unit (ReLU) in a decoder/encoder geometry as the generating function.  They proposed an improved form of the deep deconvolution or canonical generating functional, which we derive in this paper and call the Heisenberg Scattering Transformation (HST).  There are also connections of the HST to Heisenberg's S-matrix,\citep{chew.61,weinberg05} the Wigner-Weyl transformation,\citep{wigner.32,weyl.50} and the Mayer Cluster Expansion.\citep{uhlenbeck63} 

Note there has been a large Lagrangian detour taken from this canonical route down the road of further expanding the Mayer Cluster Expansion, based on the weakness of correlation (that is, the BBGKY hierarchy \citep{nicholson83,glinsky.bbgky.24}), and the weakness of coupling (that is, the perturbation theory of Feynman \citep{feynman.61} and others).  This has lead to the super convergence of the Mayer Cluster Expansion being reduced to only asymptotic convergence, and to the infinities that appear in the Wilson renormalization procedure, since the second series expansion has effectively multiplied the original series expansion by infinity.  The connection of the MST to the S-matrix (to second order) and to renormalization was first recognized by Glinsky.\citep{glinsky.14}

Given this background, we develop a theory of collective behavior, comprising of characterization, forecast (simulation), and control (including optimization and stabilization) of the collective.  Collective systems are an ensemble of conservatively interacting entities.  Collective systems take many different forms.  A plasma is a collective of charged particles, a fluid is a collective of molecules, an elementary field is a collective of elementary particles, a cosmos is a collective of celestial bodies, a society is a collective of individuals, an economy is a collective of economic entities (people, families, villages, countries, and companies) that engage in trade, and a novel is a collective of letters.  This paper focuses on plasmas, fluids, elementary fields, and the cosmos.  Although we will call these collective systems, they are commonly called complex systems.  Although, at the core, these systems are analytic, holomorphic, or complex, they also display simple emergent behaviors.  Since complex can be viewed as the antonym of simple, we prefer to use the word collective.  These collectives are also closed or conservative.  What one individual loses, others gain. This does not exclude the possibility of external interactions, though, and will be discussed in Sec.~\ref{system.sec}. 

Addressing the two issues with Field Theory highlighted by Dirac, inspired by Dirac's view that group symmetry is primal, and building upon the work of Gell-Mann, Heisenberg, Mallat, and genAI, we develop this canonical transformation approach to collective behavior.  In Sec.~\ref{gen.sec}, we start by assuming a group symmetry to the motion of the collective, leading to the conservation of a real function, $H(p,q)$, that is the infinitesimal generator of a Lie group, $\mathscr{H}$.  The finite generator of the canonical transformation, $S_P(q)$, is derived from the infinitesimal generator, $H(p,q)$, by approximating the solution of the Hamilton-Jacobi equation.  The motion of the collective field, $f(x)$, is then generated by a canonical generating functional, $S_p[f(x)]$, a deep deconvolution with the complex logarithm (that is, $\ln(z)=\ln|z| + \text{i} \, \arg(z)$) as the activation function.  The formula for this generating functional, the Heisenberg Scattering Transformation (HST), is derived in Sec.~\ref{theory.HST.sec}, and shown in Eq.~\eqref{hst.eqn}.  This gives the Taylor expansion of $S_P(q)$, or the functional Taylor expansion of $S_p[f(x)]$, or the S-matrix, or the Mayer Cluster expansion, or the Wigner-Weyl Transformation of the collective field, $f(x)$.  The theory is quantized in Sec.~\ref{quant.sec}.  How to fit, simulate, and control (that is, optimize and ponderomotively stabilize) the collective system is described in Sec.~\ref{sim.sec}.  Section~\ref{system.sec} shows how to couple one collective system to another collective system forming systems-of-systems.  The practical computer implementation of the HST is outlined in Sec.~\ref{code_hst.sec}.  Finally, there is a discussion of the results and conclusions in Sec.~\ref{conclusions.sec}.

\section{Canonical Generating Functional Approach}
\label{gen.sec}
This section will present the new generative approach to collective behavior as a logical mathematical theory built upon a foundation of primal assumptions.  We assume that the collective behavior is governed by a Lie group, $\mathscr{H}$, symmetry.  Furthermore, this Lie group symmetry can be extended to a complex Lie group, $\mathbb{H}= \mathscr{H} \otimes \text{Ad}(\mathscr{H})$, symmetry.  The Lie group has a real function, $H(p,q)$, that is both conserved during the evolution of the collective, and is the infinitesimal group generator of the evolution of the collective.  This real function can be analytically continued to form the analytic function $H(\beta)$, which is the infinitesimal group generator of a complex Lie group $\mathbb{H}$.  This analytic function defines the topology of the complex plane $\mathbb{C}$ or Riemann surface or manifold.\citep{nehari12}  In fact, knowing the analytic function is equivalent to knowing the topology of the manifold, that is the homology classes, $\beta^*$.  These singularities can also be characterized by the Chern-Simmons three-form $\lambda \wedge d\lambda$.  Integration of this form gives the topological indexes, homology classes, or Riemann Moduli.  For 3D plasmas this is the magnetic helicity,\citep{glinsky.19} and for 2D fluids it is the vorticity.  The homology classes are also known as ground states,\citep{weinberg05} force-free states, Taylor relaxed states,\citep{taylor86} BGK modes,\citep{bernstein57} and emergent behaviors.\citep{gros15}  They are force-free states because the collective system is free from the influence of external systems, that is external forces, due to the effective infinite mass of the collective system at these points since its frequency is zero there.  

\begin{figure}
\noindent\includegraphics[width=\columnwidth]{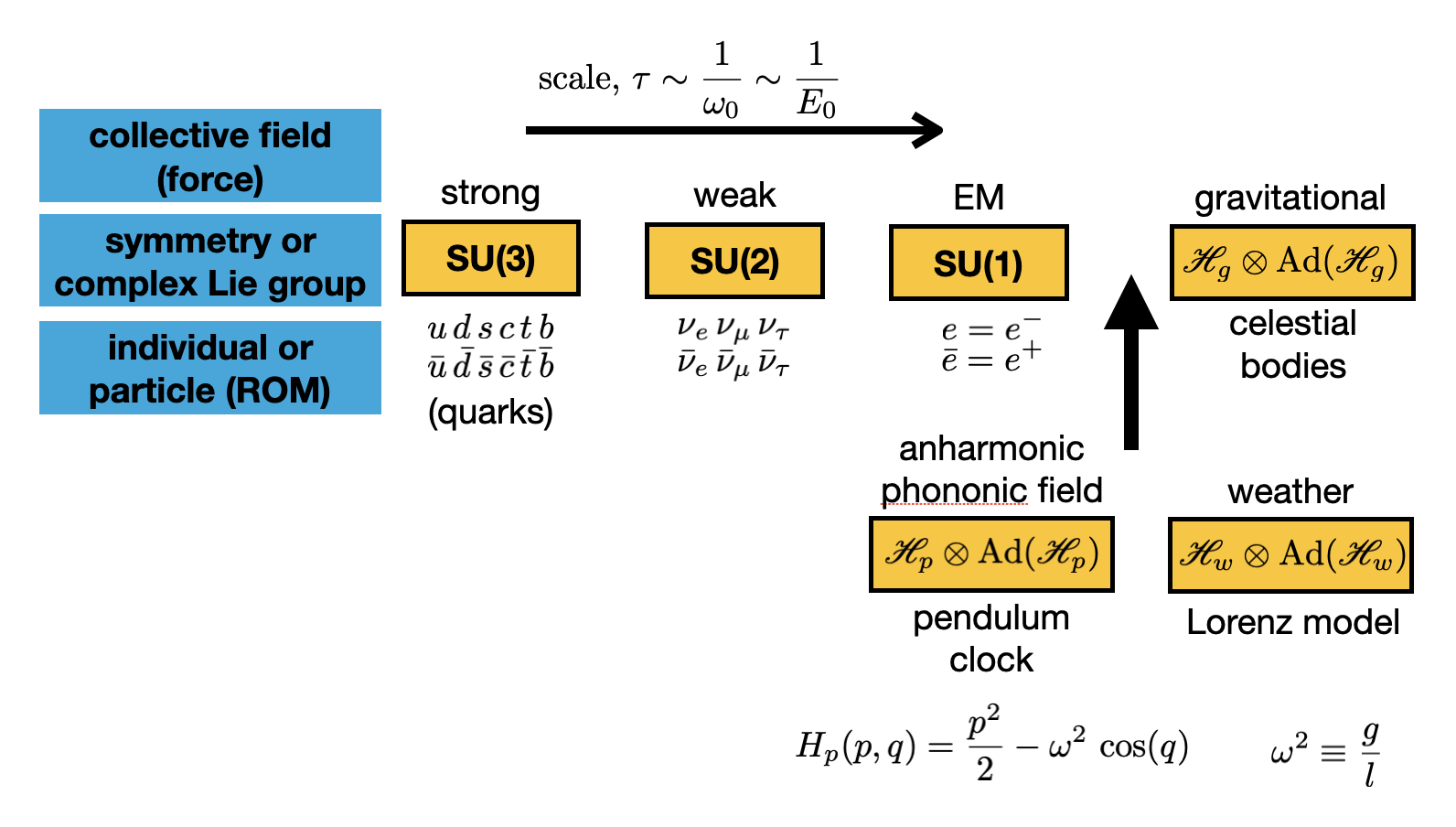}
\caption{\label{lie.groups.fig} Complex Lie group symmetries and the four corresponding elemental forces.  Shown, as a function of increasing scale $\tau$, for each case, are the force or collective field, the symmetry or complex Lie group, and the individuals or particles.  Also shown are an anharmonic phononic field based on the Hamiltonian $H_p(p,q) = p^2/2 - \omega^2 \, \cos(q)$, and the weather field.  The characteristic frequency scale is $\omega_0$, and the characteristic energy scale is $E_0$.}
\end{figure}

The four elemental forces are associated with the four elemental symmetries, shown in Fig.~\ref{lie.groups.fig}.  In order of increasing scale, they are:  SU(3) for the strong force, SU(2) for the weak force, SU(1) for the EM force, and $\mathbb{H}_g$ for the gravitational force.  The same is true, given any Hamiltonian function, $H(p,q)$.  For a pendulum clock with the Hamiltonian $H_p(p,q)=p^2/2-\omega^2 \, \cos(q)$, one can form the complex Lie group $\mathbb{H}_p=\mathscr{H}_p \otimes \text{Ad}(\mathscr{H}_p)$.  For the weather with the Hamiltonian $H_w(p,q)$, one can form the complex Lie group $\mathbb{H}_w=\mathscr{H}_w \otimes \text{Ad}(\mathscr{H}_w)$.  A well known three dimensional weather model is the Lorenz model.\citep{lorenz63}  Both of these collective systems are intermediate in scale between the EM force and the gravitational force.

\begin{figure}
\noindent\includegraphics[width=12pc]{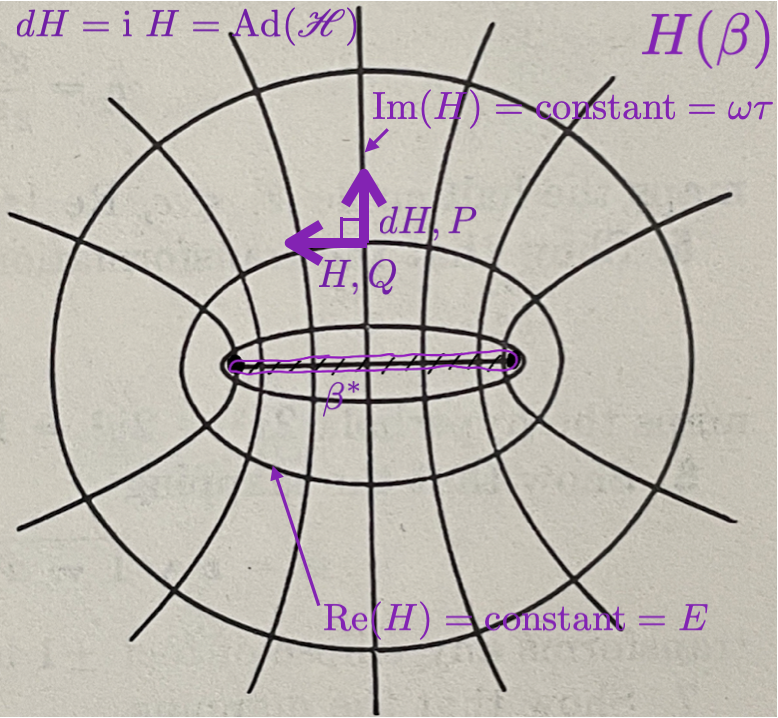}
\caption{\label{analytic.function.fig} Plots of the level sets of the function $H(\beta)=(\beta+1/\beta)/2=(\beta+\text{i})(\beta-\text{i})/2\beta$ showing the contours of $\text{Re}(H(\beta))=\text{constant}=E$ and $\text{Im}(H(\beta))=\text{constant}=\omega\tau$.  Equivalently, the group actions of $H$, in the $Q$ direction, and $dH$, in the $P$ direction.  Shown also are the branch points at $+1$ and $-1$, as black dots, and the branch cut between those points, as the hashed line.  This is a two-sheeted Riemann surface (that is manifold) with sheets that cover the entire plane and are connected by the branch cut.  The equivalence class or homology class or ground state, $\beta^*$, is the cycle around the branch cut.  Specification of the branch cut, $\beta^*$, is equivalent to specifying $H(\beta)$, since $H(\beta)$ is the solution to Laplace's equation given $\beta^*$.  This mapping is of importance in certain aerodynamical applications since it maps certain circles to Joukowski profiles which have the general character of airplane wing profiles.}
\end{figure}

\begin{figure}
\noindent\includegraphics[width=\columnwidth]{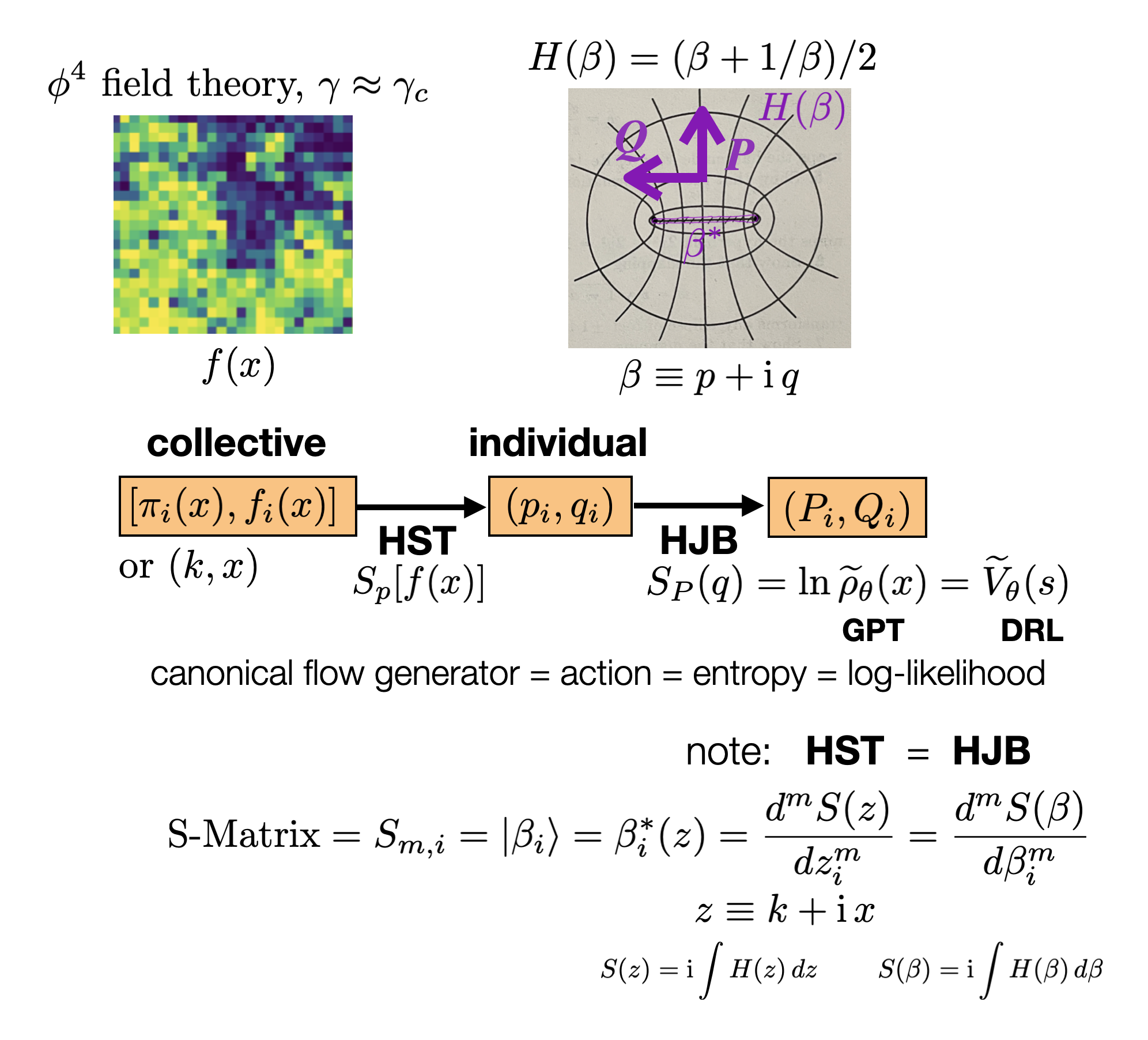}
\caption{\label{hst.gen.fig} Graphic of the generating functional (HST), $S_P[f(x)]$, and the generating function (HJB), $S_p(q)$, of the canonical transformation from the collective fields and field momentums, $[\pi_i(x), f_i(x)]$, to the ROM, $(p_i,q_i)$, to the solution of the HJB equation, $(P_i,Q_i)$.  Shown is an example of a collective $\phi^4$ field near its critical point, $\gamma=\gamma_c$, where there are long range correlations, and an example of an analytic function, $H(\beta)=(\beta+1/\beta)/2$, in the ROM space, $\beta=p+\text{i} \, q$, with the $Q$ and $P$ mutually orthogonal foliations shown.  The relationships of the generating function $S_p(q)$ to the approximate value function of DRLs, $\widetilde{V}_\theta(s)$, and the approximate log-likelihood of GPTs, $\ln \widetilde{\rho}_\theta(x)$, are shown.  The equivalence of the functional Taylor expansion of $S_P[f(x)]$ to the Taylor expansion of $S_p(q)$, is also shown.}
\end{figure}

The function $H(\beta)$ gives two foliations of the manifold.  The first are leaves where $\text{Re}(H(\beta))=\text{constant}$ and an orthogonal set of leaves where $\text{Im}(H(\beta))=\text{constant}$.  This is shown in Fig.~\ref{analytic.function.fig}.  The finite generating function of the canonical transformation, $S_P(q)$, and the energy, $E(P)$, are the solution of the Hamilton-Jacobi-Bellman (HJB) equation, 
\begin{equation}
\label{hjb.eqn}
        \boxed{\frac{\partial S(P,E;q,\tau)}{\partial \tau} + H(\partial S / \partial q,q) \textcolor{red}{- \nu \, S} = 0,}
\end{equation}
where $q$ is the canonical coordinate, $p$ is the conjugate canonical momentum, $P$ is the new conjugate  canonical momentum, $\tau$ is the metric on $x$ (e.g., $\tau(x,y,z,t)=\sqrt{t^2-(x^2+y^2+z^2)/c^2}$ for a hyperbolic Minkowski metric on $(t,x,y,z)$-space), $\nu$ is a coefficient of resistivity (introduced by Bellman for resistive control, and solvability of the HJB equation) and  
\begin{equation}
\label{action.eqn}
    S(P,E;q,\tau) \equiv S_P(q)- E(P) \tau
\end{equation}
is the action on extended phase space.  The finite generating function is also called the action, entropy, or the log-likelihood.  The generating function, that is a solution to Eq.~\eqref{hjb.eqn}, generates a canonical transformation from $(p,q) \to (P,Q)$, where the motion is cyclic in $Q$, that is linear motion.  Bellman has been added because of his recognition of the importance of the canonical generating function to optimal control theory,\citep{kalman63} which he called the approximate Value Function instead of the action, the state $s$ instead of the canonical coordinate $q$, and the approximation parameter $\theta$, instead of the new conjugate momentum $P$.  The reason for the word approximate being used is that the Value Function is solved for by using what is called Feature Value Iteration, where the features are used to approximate the Value Function, $\widetilde{V}_\theta(s)$, and the $\theta$ are the parameters of the approximation.  The problem with this approach is that the parameters do not approximate the solution, they are the solution.  The action generates a canonical transformation from $(p,q)$ to $(P,Q)$, as shown in Fig.~\ref{hst.gen.fig}.  The two foliations are also defined by $P=\text{constant}$ and $Q=\text{constant}$, as shown in Fig.~\ref{analytic.function.fig}.

\begin{figure}
\noindent\includegraphics[width=\columnwidth]{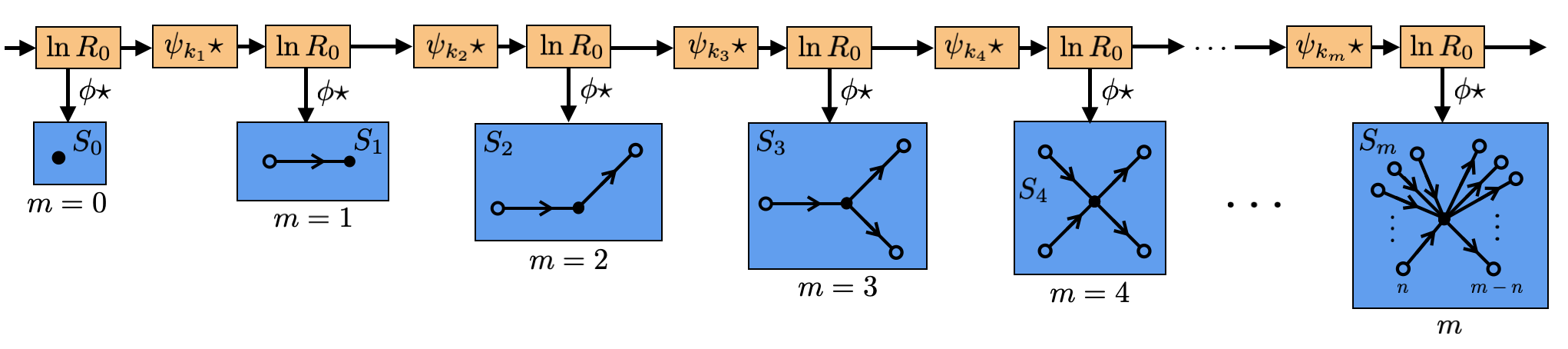}
\caption{\label{hst.fig} Diagrammatic representation of the Heisenberg Scattering Transformation $S_p[f(x)]$  (HST), as a Mayer Cluster Expansion or $m$-body Scattering Cross Sections.  Note that i's have been omitted for clarity.}
\end{figure}

The finite generating functional form of the action, $S_p[f(x)]$, is the Heisenberg Scattering Transformation (HST).  The form of this transformation is derived in Sec.~\ref{theory.HST.sec} and appears as Eq.~\eqref{hst.eqn}.  A graphical representation of the HST expansion is shown in Fig.~\ref{hst.fig}.  This canonical generating functional generates a canonical transformation from the canonical field momentum and canonical field to the canonical momentum and canonical coordinate, that is $[\pi(x),f(x)] \to (p,q)$.  It satisfies the functional form of the Hamilton-Jacobi equation,
\begin{equation}
\label{hjb.functional.eqn}
    \frac{\partial S[p,E;f(x),\tau]}{\partial \tau} + H[\pi=\delta S / \delta f,f] = 0,
\end{equation}
so that the motion will be linearized, that is constrained to a low dimensional linear subspace.  Note that the functional form of the Hamilton-Jacobi Equation, shown in Eq.~\ref{hjb.functional.eqn}, has been shown to be equivalent to Schrodinger's Equation.\citep{HJ.Schrodinger.26}

These expansion coefficients can be viewed several ways.  They can be viewed as the Mayer Cluster Expansion in the order of the correlation.  For the Mayer Cluster Expansion, $S_1$ tells how one body is distributed, $S_2$ tells how two bodies are correlated, $S_3$ tells how three bodies are correlated, $S_4$ tells how four bodies are correlated, and $S_m$ tells how $m$ bodies are correlated.  They can be viewed as the S-matrix or the $m$-body Scattering Cross Section expansion in the order of the scattering.  For Heisenberg's S-matrix, $S_1$ is the one body scattering cross section, $S_2$ is the two body scattering cross section, $S_3$ is the three body scattering cross section, $S_4$ is the four body scattering cross section, and $S_m$ is the $m$-body scattering cross section.  They also can be viewed as the $m$-body Green's Function expansion, and the $m^\text{th}$-order complex curvatures of the manifold.  

The functional Taylor expansion coefficients of the HST, $S_p[f(x)]$, are equal to the Taylor expansion coefficients of $S_P(q)$.  The results of a Principal Components Analysis (PCA) on the output of the HST, $\left| \beta_i(z) \right>$, where there is one component for each field, are the functional Taylor expansion coefficients of the HST or the S-matrix, $S_m$. They are
\begin{equation}
\label{smatrix.eqn}
\begin{split}
    \frac{\delta^m S[f(x)]}{\delta f^m} &= \frac{d^m S(z(x))}{dz^m} = S_m \\
    &=\left| \beta(z) \right> = \frac{d^m S(\beta)}{d\beta^m} = \frac{d^m S(q)}{dq^m},
\end{split}
\end{equation}
where $z \Rightarrow \beta$ as $\text{HST} \Rightarrow \text{HJB}$, $S(z)=\text{i} \int{H(z) \, dz}$, and $S(\beta)=\text{i}\int{H(\beta) \, d\beta}$.  Note that $z \equiv k + \text{i} \, x$ and $\beta \equiv p + \text{i} \, q$ so that $\beta^*_i(z)$ gives the transformation $(k,x) \to (p,q), \text{ that is } [\pi(x),f(x)] \, \to (p,q)$.  These relationships are shown in Fig.~\ref{hst.gen.fig}.  More mathematical details can be found in Glinsky and Sievert.\citep{glinsky23b}

Another way of understanding this equivalency of the individual and the collective is that the simple singularities, $\beta^*_i$, defining the topology of the individual are deeply convolved into the singularity spectrums, $\left| \beta_i(z) \right>$ or $\beta^*_i(z)$, of the collective.  Note that these are localized (that is, a function of position $x$) complex spectrums that are a function of a complex variable, $z \equiv k + \text{i} \, x$.

There is nothing special about using the Hamiltonian to generate the Weyl-Heisenberg group $\mathbb{H}^1$.  The procedure can be repeated with any other group symmetry of the dynamics $\mathbb{G}^{n_j}_j$.  Form the analytic function $G(\beta)$, and use it to further foliate the manifold.  This can continue up to the point that
\begin{equation}
    n = 1+ \sum_{j=1}^{N_g}{n_j},
\end{equation}
when the motion becomes integrable, that is fully foliated.  Here, $n$ is the number of fields, $n_j$ is the number of group parameters in group $j$, and $N_g$ is the number of groups.  The dynamics is now governed by the product group
\begin{equation}
    \mathbb{H}^1 \otimes \mathbb{G}^{n_1}_1 \otimes \mathbb{G}^{n_2}_2 \otimes \dots \otimes \mathbb{G}^{n_j}_j \otimes \dots \otimes \mathbb{G}^{n_{N_g}}_{N_g}.
\end{equation}
This leads, via Noether's Theorem, to additional constants of the motion for each additional symmetry, and additional quantum numbers when the theory is quantized in Sec.~\ref{quant.sec}.

In summary, the collective follows a geodesic motion determined by the group symmetries of the collective.   Given the topology of the complex Lie group, $\beta^*$, the geodesics can be found by solving Laplace's equation with the boundary conditions $\beta^*$.  It is a question of the geometry of the physics,\citep{frankel.11} and discovering the topology of the collective's evolution.

\section{Theory of the Heisenberg Scattering Transformation (HST)}
\label{theory.HST.sec}
\begin{figure}
\noindent\includegraphics[width=\columnwidth]{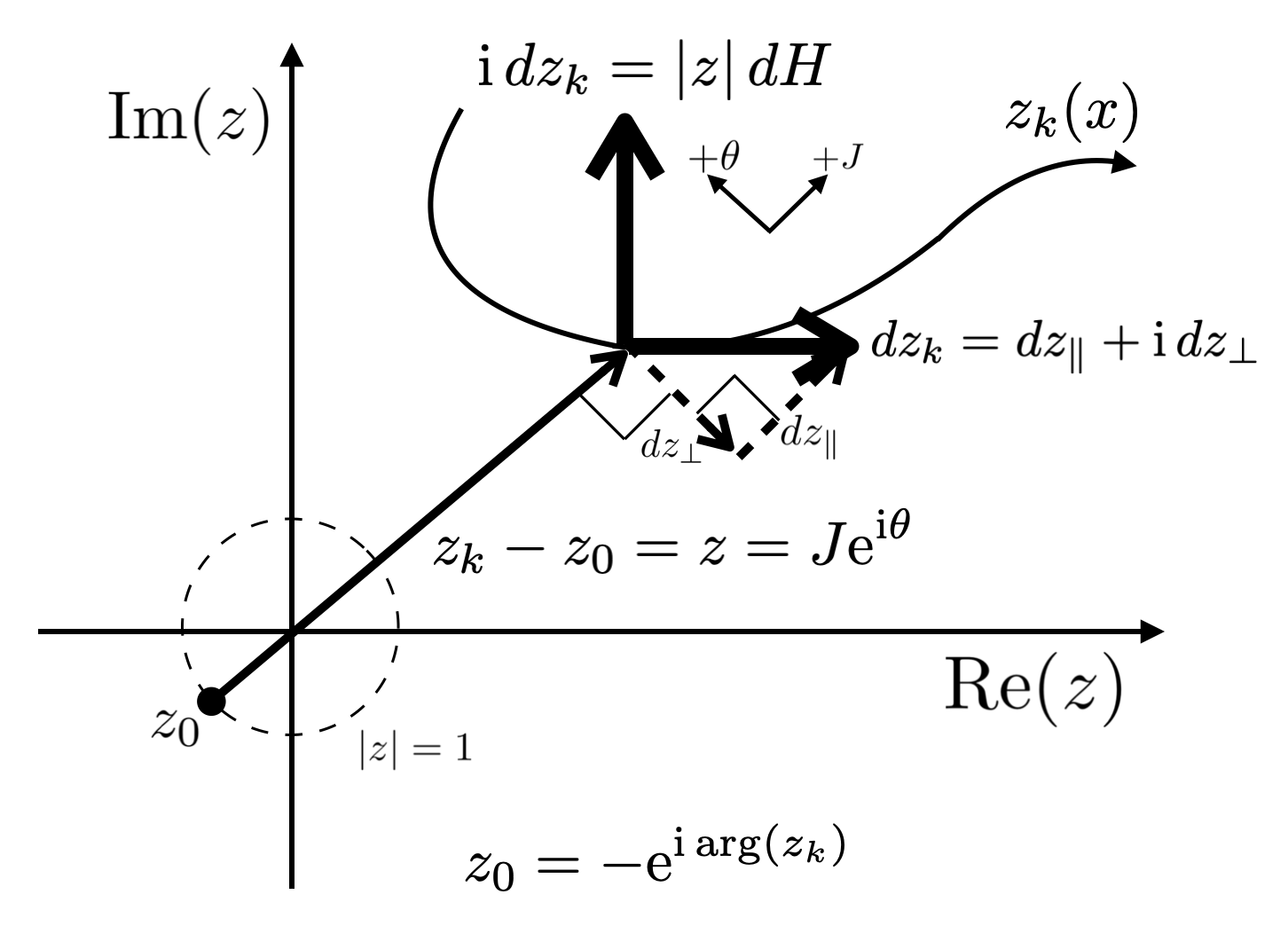}
\caption{\label{hst.diagram.fig} Trajectory in $\mathbb{C}$ given by $z_k(x)$ projected onto polar coordinates about $z_0$ and rotated to the normal ($dH$ covector) direction.  See Eq.~\eqref{lg.gen.eqn} for details.}
\end{figure}

The transformation we endeavor to construct takes the dynamics to a manifold that is a linear subspace of a Hilbert space with basis vectors that are the solution to the Renormalization Group Equations (RGEs).  In this subspace, the dynamics is geodesic motion with the topology given by the Hamiltonian function $H(z)$, an analytic function on $\mathbb{C}^n$, where $n$ is the number of fields.  Another way of looking at this is that the motion is harmonic or holomorphic.  We start by constructing a logarithmic generator of the function $\bar{H} \equiv \text{i} H$.  First, Taylor expand the function $\bar{H}$ about $z_0$ giving
\begin{equation}
    \bar{H}(z) = \sum_{m=0}^{\infty}{\frac{1}{m!} H_m(z_0) \, (z-z_0)^m},
\end{equation}
where 
\begin{equation}
    H_m(z) = \frac{d^m \bar{H}(z)}{d z^m} = \frac{dS_m(z)}{dz},
\end{equation}
which serves as a definition of $S_m(z)$.  Now, we need to find an expression for $S_m$ given $S_{m-1}$.  Start by assuming that $S_m$ is a functional of the field and the field momentum or the complex field, that is
\begin{equation}
    S_m(z(x)) = F[\pi(x), f(x)] = F[f(x)=\pi(x) + \text{i} \, f(x)].
\end{equation}
Define the wavelet transformation
\begin{equation}
    \psi_k \star f(x) = \int{\psi_k(x') \, f(x-x') \, dx'},
\end{equation}
where $\psi_k(x)$ are a normalized, orthogonal, localized and harmonic (that is coherent states) such that
\begin{equation}
    \psi_k(x) \equiv k^2 \,\psi(kx),
\end{equation}
and
\begin{equation}
    \phi_{kx}(x') \equiv k^2 \, \phi(k(x'-x)),
\end{equation}
where $\psi(x)$ and $\phi(x)$ are the Mother and Father wavelets that satisfy the Littlewood-Pauley condition.  Consider the path
\begin{equation}
    z_k(x) \equiv \psi_k \star S_{m-1}(z(x)).
\end{equation}
It can be proven, using the fact that $\pi(x)$, the conjugate field momentum, and $f(x)$ satisfy the field equations (effectively Hamilton's equations), $\psi_k(x)$ are coherent states, and the functional chain rule; that $z_k(x)$ is a trajectory on the complex plane that satisfies the Cauchy-Riemann conditions.  That is to say, $z_k(x)$ is an analytic trajectory.  We now want to calculate the analytic function that gives this trajectory.  See Fig.~\ref{hst.diagram.fig} for a graphical representation of this derivation.  Take the covector $dz_k$ along this trajectory, change to radial coordinates about $z_0=-\text{e}^{\text{i} \arg(z_k)}$ such that $z=z_k-z_0=J \text{e}^{\text{i} \theta}$, and rotate by $\pi/2$ (that is multiply by $\text{i}$), to get these expressions for the covector
\begin{equation}
\label{dS.eqn}
\begin{split}
    dS_m(z) &= \text{i} \frac{dz_k}{|z|} = \text{i} \left( \frac{dz_\parallel}{|z|} + \text{i} \frac{dz_\perp}{|z|} \right) \\
    &= \text{i} \left( \frac{dJ}{J} + \text{i} \theta \right) = \text{i} \, d(\ln(J \text{e}^{\text{i} \theta})) \\
    &= \text{i} \, d(\ln z) = d(\text{i} \, \ln z).
\end{split}
\end{equation}
Defining
\begin{equation}
\label{R0.eqn}
    R_0(z) \equiv \frac{1}{\text{i}} h^{-1}(2z/\pi),
\end{equation}
and $h(z) \equiv (z+1/z)/2$, one gets the recursion relation
\begin{equation}
\label{lg.gen.eqn}
\begin{split}
    S_m &= \text{i} \, \ln z = \text{i} \, \ln(R_0(z_k)) \\
    &= \text{i} \, \ln R_0 \, \psi_{k_m} \star S_{m-1}.
\end{split}
\end{equation}
Remember that the complex logarithm is
\begin{equation}
    \ln(z) = \ln|z| + \text{i} \, \arg{(z)}.
\end{equation}
With the definition of $R_0$ given in Eq.~\eqref{R0.eqn},
\begin{equation}
    \text{i} \, \ln(R_0(z)) \xrightarrow[\epsilon \to 0, \, \text{for }z \in (-\pi/2+\text{i} \epsilon, \pi/2+\text{i}\epsilon)]{} z 
\end{equation}
is a compact mapping for $z\in (-\pi/2,\pi/2)$.  We use $f(x)$ as shorthand for any $F[\pi(x),f(x)]$.  Start with
\begin{equation}
    S_0 = \text{i} \, \ln R_0 \, f(x).
\end{equation}
Remove the explicit $x$ dependence with a final convolution, $\phi_{kx} \star$.  Choose normally ordered paths,
\begin{equation}
\label{wick.eqn}
    k_{m+1} < k_m \text{ and } x_{m+1} > x_m,
\end{equation}
so that we can define
\begin{equation}
    k = \sum_{m=1}^\infty{k_m} \text{ and } x = \sum_{m=1}^\infty{x_m}.
\end{equation}
This yields the following expression for the iterative logarithmic generating functional (the HST)
\begin{equation}
\label{hst.eqn}
    \boxed{S_m[f(x)](z) = \phi_{kx} \star \left( \prod_{n=1}^{m}{\text{i} \ln R_0 \psi_{k_n} \star} \right) \text{i} \ln R_0 f(x),}
\end{equation}
where $z \equiv k + \text{i} \, x$.  This Logarithmic Generating functional (LG) generates, by construction, $S(z)= \text{i}\int{ H(z) \, dz}$.  It is not the exponential generator, propagator, or partition function, given by $U=\text{e}^{-(\text{i}/\hbar) \int{H \, d\tau}}$ or $Z=\text{e}^{\text{i} S / \hbar}$, as is done in the Lagrangian approach.  It follows that
\begin{equation}
    \text{LG}''(z) = 0,
\end{equation}
which means that the transformation has flattened the space so that the infinitesimal generator, $\text{i}H$, is now the finite group generator.  We have transformed to the basis $\left|\beta(z)\right>$ where $\beta \equiv p + \text{i} \, q$, the Hilbert space of all possible solutions to the RGEs, corresponding to $S_m=\int{H_m \, dz}$, the S-matrix.  The motion will be confined to an $n$-dimensional complex linear hyperplane $\mathbb{C}^n$, where $n$ is the number of fields, and will be geodesic motion on that submanifold, determined by the analytic function $H(\beta)$, with $m^\text{th}$-order complex curvature given by the S-matrix,
\begin{equation}
    \frac{d^{m-1} H(\beta)}{d\beta^{m-1}} = \frac{d S_{m-1}}{dz} = S_m \sim \frac{d(\ln z)}{dz} \sim \frac{1}{\beta_{\text{rc},m}},
\end{equation}
where $\beta_{\text{rc},m}$ is the $m^\text{th}$-order complex radius of curvature, and we used the expression $S_{m-1} \sim \ln z$ given by Eq.~\eqref{dS.eqn}.  Streamlines of the motion generated by $H$ (the $\mathscr{H}$ group action) are given by $\text{Re}(H(\beta))=\text{constant}$, and streamlines of $dH$ (the $\text{Ad}(\mathscr{H})$ group action) are given by $\text{Im}(H(\beta))=\text{constant}$.

\begin{figure}
\noindent\includegraphics[width=\columnwidth]{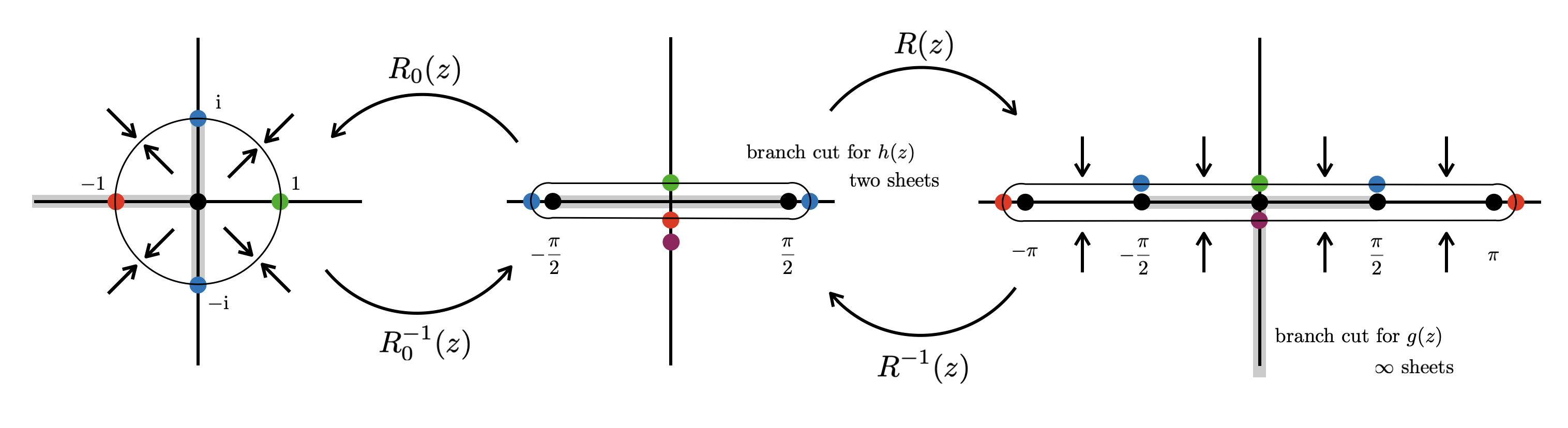}
\caption{\label{analytic.rectifier.fig} The conformal mappings, $R_0(z)$ and $R(z)$.  Branch points are shown as the black dots, and branch cuts are shown as the thin gray rectangles.  Reference points on the contours are shown as colored dots.  Black arrows indicate convergence of the compact mapping, $\ln(z)$.  Note that the part of the contour between $-\pi/2+\text{i}\epsilon$ and $\pi/2+\text{i}\epsilon$ is mapped to itself, by the analytic rectifier function, $R(z)$. The result is a compact conformal mapping converging to the fixed contour of a unit helix unrolled around the real axis, with repeated application of $R(z)$ and bination.}
\end{figure}

\begin{figure}
\noindent\includegraphics[width=\columnwidth]{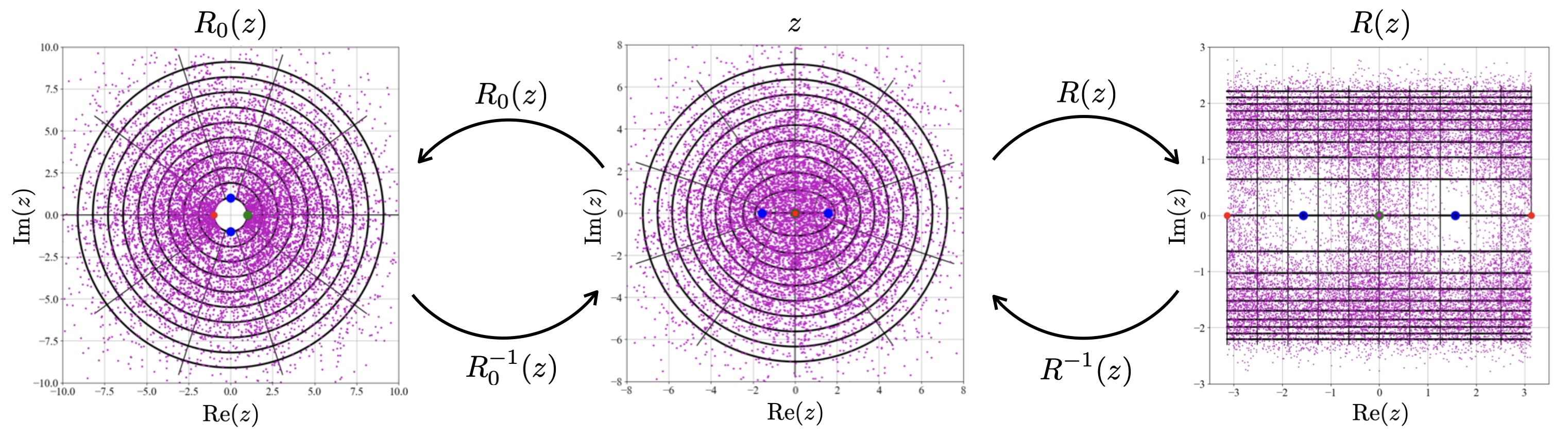}
\caption{\label{log.linearizer.fig} The conformal mappings, $R_0(z)$ and $R(z)$.  Reference points are shown as large colored dots.  A Gaussian distribution of points in the original domain is shown as small purple dots.  Reference orthogonal coordinate lines are shown as thin black lines.  Note that the $R_0(z)$ function transforms to polar coordinates, and the $R(z)$ function transforms to cartesian coordinates.}
\end{figure}

We take a closer look at the analytic rectifier function, $R(z) \equiv \text{i} \, \ln(R_0(z))$.  Both the $R(z)$ and $R_0(z)$ conformal mappings are shown in Fig.~\ref{analytic.rectifier.fig}.  Note that the $R_0(z)$ mapping takes the contour around the branch cut between $-\pi/2$ and $\pi/2$ to a contour around the unit circle.  There are two sheets to the Riemann surface of $R_0$.  The first sheet is mapped to the area outside the unit circle, the other sheet is mapped to the area inside the unit circle.  The complete conformal $R(z)$ mapping takes the contour twice circulating around the branch cut between $-\pi/2$ and $\pi/2$, through the branch cut for $\exp(z)$, to the contour that circulates once around $-\pi$ and $\pi$.  The part of the contour between $-\pi/2+\text{i}\epsilon$ and $\pi/2+\text{i}\epsilon$ is mapped to itself.  Remember that the Riemann surface for $\exp(z)$ has an infinite number of sheets.  The mapping is compact, converging to the real axis.  The result is a compact conformal mapping converging to the fixed contour of a unit helix unrolled around the real axis, with repeated application and bination.  The analytic rectifying function, $R(z)$, can also be identified as the well-known analytic Green’s function, $G(z)=R(z)$ (see \citet{nehari12} Eq. 42 on page 181).  Another way of looking at the $R(z)$ function is as a nonlinear function that linearizes a complex nonlinear distribution or coordinate system, as shown in Fig. \ref{log.linearizer.fig}.  The $R_0(z)$ function transforms to polar coordinates, while the $R(z)$ function transforms to cartesian coordinates.  This is why the motion is confined to a an $n$-dimensional complex linear hyperplane, $\mathbb{C}^n$, after the HST.

Now lets review what the HST is doing. The $\psi_k \star$ is generating $z_k(x)$ a parametric trajectory of the analytic function $\text{i} H$.  The $\ln R_0$ conformal (canonical) transformation is flattening the space onto the cylinder by transforming into polar coordinates about $R_0$.  The $\text{i}$ conformal transformation is rotating $\pi /2$ from the tangent direction to the gradient (covector) direction.  Another perspective views it as a canonical transformation that is exchanging the field momentum with the field.  This can be recognized by the fact that $\pi(x)+\text{i} \, f(x)$ and $f(x)+\text{i} \, \pi(x)$ are canonically conjugate fields.  The choice of $R_0$ ensures that the repeated application of the mapping converges to a circle of radius $\pi$ about the origin on the cylinder.  This is because $R_0$ is a fixed point of $\ln$, that is $\ln(R_0(z)) \to z$ as $|z| \to 0$.  The convergence to the origin is exponential for large deviations, then logarithmic for small deviations.

It should be noted that the transformation is no longer stationary since the Father Wavelet only averages over as large of a patch as it has to do.  Nothing prevents this partition of unity from being summed over a larger domain in $x$, if the process is stationary over that domain.  This transformation is complex from the beginning to the end.  The connection to previous work can be seen by examining the limiting behavior of the $\ln(R_0(z))$ mapping
\begin{equation}
\begin{split}
    \ln(R_0(z)) &\xrightarrow[z \to (-\pi/2,\pi/2)]{} z/\text{i} \\
           &\xrightarrow[|z| \to \infty]{} \ln|z| \quad \text{(Glinsky and Maupin)}\\
           &\xrightarrow[|z| \to \infty]{} \ln|z| \, \text{e}^{\text{i} \, \arg(z) / \ln|z|}\\
           &\sim \sum_{k=0}^n{|z| \, \text{e}^{\text{i} \, k \, \arg{(z)}}} \quad \text{(that is, WPH)}\\
           &\xrightarrow[n = 0]{} |z| \quad \text{(that is, MST)}.
\end{split}
\end{equation}
The characteristic of the fixed point at the origin is the statement of the first limit. The second limit is effectively what has been used in the previous work of \citet{glinsky23}.  In the third limit (which keeps the small $\arg$ imaginary term), the $\ln$ chirps the pulse, generating the harmonics that are explicitly generated in the fourth line by Wavelet Phase Harmonics (WPH).\citep{mallat2020phase,zhang2021maximum,allys2020new,regaldo23}  WPH is the attempt by St{\'e}phane Mallat and his collaborators to incorporate phase into the transformation.  The original Mallat Scattering Transformation (MST) \citep{mallat.12} is just using the first term in that series.

The process of one iteration of the HST can be viewed in this simple way.  The convolution generates an analytic trajectory on the complex plane.  The ln changes to polar coordinates after translating the origin to the fixed point of the mapping, all canonical and analytic transformations.  This flattens the space, and exposes the curvature of the manifold as the expansion coefficients or S-matrix, $S_m$.  This process is simply taking the canonical derivative, $\text{i} \, d(\ln)$, where the exterior derivative is taken by $\psi_k \star$.

This Mayer Cluster Expansion, in the order of the correlation $m$, is super convergent, that is converges much faster than a geometric series.  In fact, terms of third and higher order are nearly zero.  Write the expansion as
\begin{equation}
    S_p[f(x)] = \sum_{m=0}^\infty{S_m},
\end{equation}
The mistake that has been made is to expand in the weakness of correlation (that is, the BBGKY hierarchy of plasma physics), or the strength of the coupling constant (that is, the perturbation expansion of Field theory).  Each term of the Mayer Cluster Expansion is expanded in the correlation parameter or coupling constant, $\Gamma$, 
\begin{equation}
    S_m = \sum_{n=0}^\infty{a_{mn}\Gamma^n},
\end{equation}
then the terms of like order in $\Gamma$ are collected
\begin{equation}
    A_n = \sum_{m=0}^\infty{a_{mn}},
\end{equation}
so that
\begin{equation}
    S_p[f(x)] = \sum_{n=0}^\infty{A_n \Gamma^n}.
\end{equation}
The problem is that the convergence in $A_n \Gamma^n$ is only asymptotic, so that
\begin{equation}
    A_n \Gamma^n \xrightarrow[n \to \infty]{} \infty.
\end{equation}
This is the origin of the infinities in Wilson renormalization.  Furthermore, for many cases (the most interesting cases) $\Gamma \gtrsim 1$.  To cure these two problems, Wilson and t'Hooft \citep{thooft.73} developed the mathematically illogical process of renormalization which ``solves the Renormalization Group Equations'' for $S_m$.  It essentially reverses the perturbation expansion, that is $A_n \to S_m$.

Let us be precise about how the HST ``solves the RGEs''.  The RGEs can be written as
\begin{equation}
\label{rge.eqn}
    \frac{d(\ln Z(k))}{d(\ln k)}=-C(k),
\end{equation}
where $k$ is the inverse scale, $Z_k(S)=\text{e}^{\text{i}S/\hbar}$ is the partition functional, and $C(k)$ is the scale coupling function.  Identifying the canonical derivative as
\begin{equation}
    \text{i} \, d(\ln) \sim \text{i} \, \psi_k \star \ln R_0 \sim \frac{\text{i} \, d(\ln)}{d(\ln k)},
\end{equation}
how the HST is finding the solution to the RGEs, that is Eq.~\eqref{rge.eqn}, is obvious.  Equation~\eqref{rge.eqn} is the ODE whose solution is $S(k), \text{ that is } \beta(z)$.  In summary, the HST is based on a canonical (that is, Heisenberg) approach, in contrast to the more common Lagrangian (that is, Schrodinger or Feynman path integral perturbation expansion of $\text{e}^{\, (\text{i}/\hbar) \int \lambda^1}=\text{e}^{\text{i}S/\hbar}$, where $\lambda^1$ is the Poincaré one-form or Lagrangian) approach with the need for a Wilson renormalization.

The relationship to the Werner Heisenberg’s 1943 S-matrix ideas comes from the fact that the S-matrix is the curvature, $S_m=d^{m-1}H/d\beta^{m-1}$, so we are calculating the expansion coefficients of $H(\beta)$ by calculating the S-matrix. Just take the derivative, $H_m = dS_m/dz$.  This is why Eq.~\eqref{hst.eqn} is called the Heisenberg Scattering Transformation.  Second, the HST is also closely related to the ideas of Eugene Wigner and Hermann Weyl, when they constructed the Wigner-Weyl transformation.  They wanted to transform from the collective field domain to $\mathbb{R}^n$, then solve an ODE.  The HST follows this basic philosophy by transforming from the collective field domain to $\mathbb{C}^n$, then solving Laplace's Equation for the minimal or geodesic surface.

The concept of the S-matrix was first introduced by Werner Heisenberg in 1943.\citep{heisenberg.43}  He lost interest in it, most likely, because he lacked concepts of analyticity.  It was picked up again by \citet{chew.55} in 1955 and was most completely developed by Lev Landau \citep{landau.59} and Richard Cutkosky \citep{cutkosky.60} by 1960.  A good survey of this work was written by Chew.\citep{chew.61}  This work was based on the concepts of analyticity and unitarity.  It further used the perturbative expansion (in terms of the strength of the field coupling constant) of an exponential generator employed by Richard Feynman in the path integral approach to Field Theory.  Furthermore, it evaluated the resulting integrals using a Fourier basis.  It had no physical basis for the analyticity.  Because of this, the theory was incomplete and did not form a well defined theoretical structure.  It was also very difficult to calculate, and was not evaluated past second order.  This line of research was abandoned in favor of the path integral formulation.

The work presented in this paper addresses these deficiencies.  First, analyticity is the key to this work and has a deep physical basis coming from the isomorphism between Hamilton's equations and the Cauchy-Riemann equations.  There is no need to enforce unitarity, since analyticity and sympletic structure lead to unitarity.  The exponential generator is replaced by a logarithmic generator so that the dynamics are constrained to a linear subspace.  A wavelet basis with local support is used in place of the Fourier basis so that the integrals do not suffer from infinities, more simply said, they are compatible with the evaluation of integrals on manifolds.   The analytic function $H(\beta)$ specifies the topology of the dynamical manifold.

This work is also closely related to the Wigner-Weyl transformation.\citep{case.08,wigner.32,weyl.50}  They wanted to transform from the collective field domain to $\mathbb{R}^{n}$, then solve an ODE.  The problems were they:  (1) used a global Fourier basis not compatible with calculation on a manifold, (2) only calculated to second order, and (3) did not include the logarithmic transform.  This resulted in complicated corrections to the commutator, divergences in the evaluation, and an incomplete transformation.  Another way of looking at this is that Wigner and Weyl constructed a transformation from the Hilbert space to $\mathbb{R}^{n}$, but found that they could not guarantee that the sympletic structure would be preserved.  This is because sympletic structure is not built into the algebra of $\mathbb{R}^{n}$, but symplectic structure is built into $\mathbb{C}^{n}$.  The  Cauchy-Riemann equations are Hamilton's equations.

The HST follows the basic philosophy of Eugene Wigner and Hermann Weyl of transforming from the collective field domain to $\mathbb{C}^{n}$, but uses an orthogonal local partition of unity based on coherent wavelet states to evaluate the integrals on the dynamic manifold, and embeds the complex logarithm in the transformation so that the dynamical manifold is aligned with $\mathbb{C}^{n}$, where $n$ is the number of fields.  The beauty of the HST is that the ODE is Laplace's equation (the Cauchy-Riemann equations) and the dynamical motion is simply geodesic motion, like for general relativity, given the topology of the dynamical manifold.  Chaotic Dynamics and Field Theory has been reduced to a matter of geometry --- the geometry of physics.\citep{frankel.11}

\section{Quantization of the Theory}
\label{quant.sec}
All of the physics theory up to this point is deterministic, that is, there is a state of the collective that has a unique evolution.  The problem with conservative collectives is that they have as much instability as stability.  A small change to the initial state of the system will evolve to a large difference in final state, if the time is larger than the characteristic time of the system, $1/\omega_0$.  Shown in Table \ref{quantum.scales.tab} are characteristic times for several well known systems that roughly correspond to the complex Lie groups displayed in Fig.~\ref{lie.groups.fig}.
\begin{table}
\caption{\label{quantum.scales.tab} Table of stochastic or quantum scales $\Delta \tau = 1/\omega_0$ and $\Delta E = \hbar \omega_0$ for several well known systems. For scales less than about $10^{-17}\text{s}$ (indicated by the dashed line), it becomes relativistically impossible to make differential measurements.}
\begin{tblr}{c l l}
\\
\textbf{Field Theory} & $1/\omega_0 \text{ (s)}$ & $\hbar \omega_0$ \\
\hline
gravity (cosmos) & $10^{15}$ & $10^{-12} \, \text{aeV}$ \\
weather & $10^{5}$ & $0.01 \, \text{aeV}$ \\
clock & $10^{0}$ & $1 \, \text{feV}$ \\
fluid & $10^{-3}$ & $1 \, \text{peV}$ \\
plasma (solar) & $10^{-7}$ & $10 \, \text{neV}$ \\
plasma (warm) & $10^{-10}$ & $10 \, \mu\text{eV}$ \\
plasma (dense hot) & $10^{-12}$ & $1 \, \text{meV}$ \\
atom (low Z) & $10^{-16}$ & $10 \, \text{eV}$ \\ \hline[dashed]
atom (high Z) & $10^{-19}$ & $10 \, \text{keV}$ \\
QED & $10^{-21}$ & $1 \, \text{MeV}$ \\
QCD & $10^{-24}$ & $1 \, \text{GeV}$ \\
\hline
\end{tblr}
\end{table}

We now turn our attention to the measurement and prediction of the state of a collective system.  Werner Heisenberg realized that it was not possible to make space-like measurements of the state of a relativistic or collective system.  It would take knowledge of the future.    His statement of this is the Heisenberg Uncertainty Principle,
\begin{equation}
    \{p,q\} = \frac{\text{i}}{\hbar} [\pi(x),f(x)] = 1.
\end{equation}
For a collective system, one needs to know the future evolution of the collective in order to know the present state of the collective.  For a relativistic system, the future state of the system can not be communicated back to the current state.  Such systems must be treated statistically, that is as stochastic systems.  Max Born built on this with the Born Rule, which states that the only way that a statistical measurement can be made on a system is by exerting a force on the system.  The problem is that the force changes the system, so that the measurement is entangled with the system.

There has been a lot of confusion in the literature about the concept of entanglement.  The origin of the confusion is the 1927 Solvay Conference, where there was a multi-day dialogue between Bohr and Einstein.\citep{salam.90}  The result was Einstein formulating the EPR paradox.\citep{epr.35}  This has given rise to the concept of Schrödinger's Cat, and the related question, ``Does a tree fall in the woods, if there is no one there to see it?''  For the case of Schrödinger's Cat, the question is, ``Is Schrödinger's Cat alive or dead if there is no one there to observe it?''  This has been simply stated as the question, ``Is there a reality?''

The key to resolving the EPR paradox is realizing that the system that is statistically measured is not that same as the system that is simply observed.  Schrödinger's Cat is either alive or dead irregardless of whether it is being observed.  There is a reality.  A non-destructive retrospective observation can be made of whether it was alive.  From these non-destructive retrospective measurements, it can be learned that the cat has been alive for five years and that cats live for about another eight years if they have been alive for five years.  From this knowledge, a prediction can be made that it is likely that the cat will live for another eight years.  An example of a destructive statistical measurement that exerts a force on the system is stabbing the cat, so that if the cat is alive, it falls to the ground and dies.  If it was already dead, nothing changes.  If the cat falls to the ground, the cat was alive.  If nothing changes, it was dead.  In either case, it now can be predicted with certainty that the cat will be dead in the future.  Any cat passing where the measurement is being made is killed.  Stabbing the cats has exerted a force on the system changing the dynamics, that is it has reduced the life span of cats.  The destructive measurement is not measuring the original reality, it is measuring an altered reality.  There is an entanglement of measurement and reality.  For the case of the tree in the woods, the destructive measurement would be to chop the tree down, so that if the the tree had not fallen to the ground, it falls to the ground.  If it had fallen to the ground, nothing changes.

When there is a transient in the system, such as one starting to stab cats that pass by to see if they are alive, the topology of the dynamical manifold will be changed by the introduction of a singularity at the location where the cats are being stabbed.  This locally pulls on the dynamical manifold, causing a local change to the geometry, that is warping of the dynamical manifold.  This launches a wave, analogous to a gravitational wave in space time, that travels at the group velocity to the location of the passive non-destructive observer of manifold curvature.  After a travel time from the location of the destructive observer, who is stabbing cats, to the passive observer of dynamical manifold curvature, there will be a steady state change to the curvature of the dynamical manifold at the location of the passive observer.  It will take the time it takes the cats to walk from the destructive observer to the passive observer before the passive observer can know that the destructive observer is stabbing the cats, reducing their performance and life times.

To quantize this theory, consider the case of observing the system at dynamical time, $\tau_0$, then predicting its state at a later time, $\tau_0+\Delta \tau$.  If $\Delta \tau$ is less than the natural periods of the system given by $2 \pi / \omega_0$, there is no problem in predicting the state of the system.  This is the classical limit.  Due to the nonlinear character of the behavior, even if the system is integrable, there will be an exponential divergence of points on the submanifold on times greater than $2 \pi / \omega_0$.  This is the limit where the system must be treated statistically.  The system will remain close to the geodesic surface defined by $\text{Re}(H(\beta))= \text{constant}$ or $P=\text{constant}$, but its position on the geodesic, that is $Q$, will not be known.  Since this phase, $Q$, is uniform and periodic, the action must be quantized to ensure periodicity in the probability distribution.  So, the energy of any state can be written as $E = E_0 + \Delta E_n$, where $E_0=\text{Re}(H(\beta^*))$, $J=nh$, $E=n \,\hbar \omega_0$, and $n$ is a rational number ($\Delta n$ is an integer for bound states).  

This is a subtle, yet significant, departure from the traditional view of quantization.  It has the ramification that there is now no difference between the stochastic uncertainty of nonlinear dynamics, quantum mechanical uncertainty, or the statistical uncertainty of kinetic theory (that is statistical and continuum mechanics).  It also enables a systematic unified framework for all dynamical systems ranging from QuantumChromoDynamics (QCD), QuantumElectroDynamics (QED), to atomic physics, to pendulum clocks, to the weather, to gravity/cosmology.  The only difference is the time and energy scales of the quantization that range from $10^{-24}$ to $10^{15}$ sec and $10^{-12} \, \text{aeV}$ to $1 \, \text{GeV}$, as shown in Table \ref{quantum.scales.tab}.

An intuitive way of looking at quantization is to consider the following situation.  You have a pendulum clock and want to predict the position of the pendulum exactly a day later.  Because the period of the pendulum is a second, you have no idea what the phase of the pendulum, $Q$, will be, or how many times it will have gone back or forth a day later.  You do know that $P$ will not have changed and the distribution must be uniform and periodic in $Q$, though.  Therefore, the action, $E(P)/\omega_0$, must be quantized to ensure periodicity in the probability distribution.

This quantization of the $\mathscr{H}$ group action, that is $E(P)/\omega_0$, is the primary quantization of the collective system, and yields the fermion particles of the collective system.  Secondary quantization is quantization of the $\text{Ad}(\mathscr{H})$ group action, that is $S_P(q)$, yielding the field bosons of the collective system.  The field bosons couple the collective system to other collective systems, that is statistical mechanics or kinetic theory, through which dissipation and statistical equilibrium with a heat bath can be introduced. This will be discussed more in Sec.~\ref{system.sec}.  The process of quantization simply comes from enforcing a periodic boundary condition on the probability in action (that is around a cycle), once the motion becomes chaotic and must be treated statistically.  The topology of the $\mathbb{H}=\mathscr{H} \otimes \text{Ad}(\mathscr{H})$ group is that of a torus, $T^2$.  Primary quantization enforces periodicity of the action around the $\mathscr{H}$ generated cycle or homology class (in the $H$ direction), and secondary quantization enforces periodicity of the action around the $\text{Ad}(\mathscr{H})$ generated cycle or homology class (in the $dH$ direction).

The difference between statistical (quantum) observation and deterministic (classical) observation is whether the time between observations is long enough to allow the system to become ergodic (in nonlinear dynamics terminology) or de-correlate (in quantum mechanics terminology).  When a system is moving at relativistic velocity, the system can never be observed classically in a space-like interval, that is a rest frame.  This is the traditional $\hbar$ uncertainty limit.  In non-relativistic systems it is possible to observe the system classically in a rest frame.  If the system is observed quantum mechanically, that is statistically, there is a natural uncertainty limit associated with the action of the ground state, not the action of relativistic motion.  Note that even for relativistic systems, a classical observation can be made in the frame of the particle, but not in a stationary space-like frame.

\section{Generative AI Simulation and control}
\label{sim.sec}
\begin{figure}
\noindent\includegraphics[width=\columnwidth]{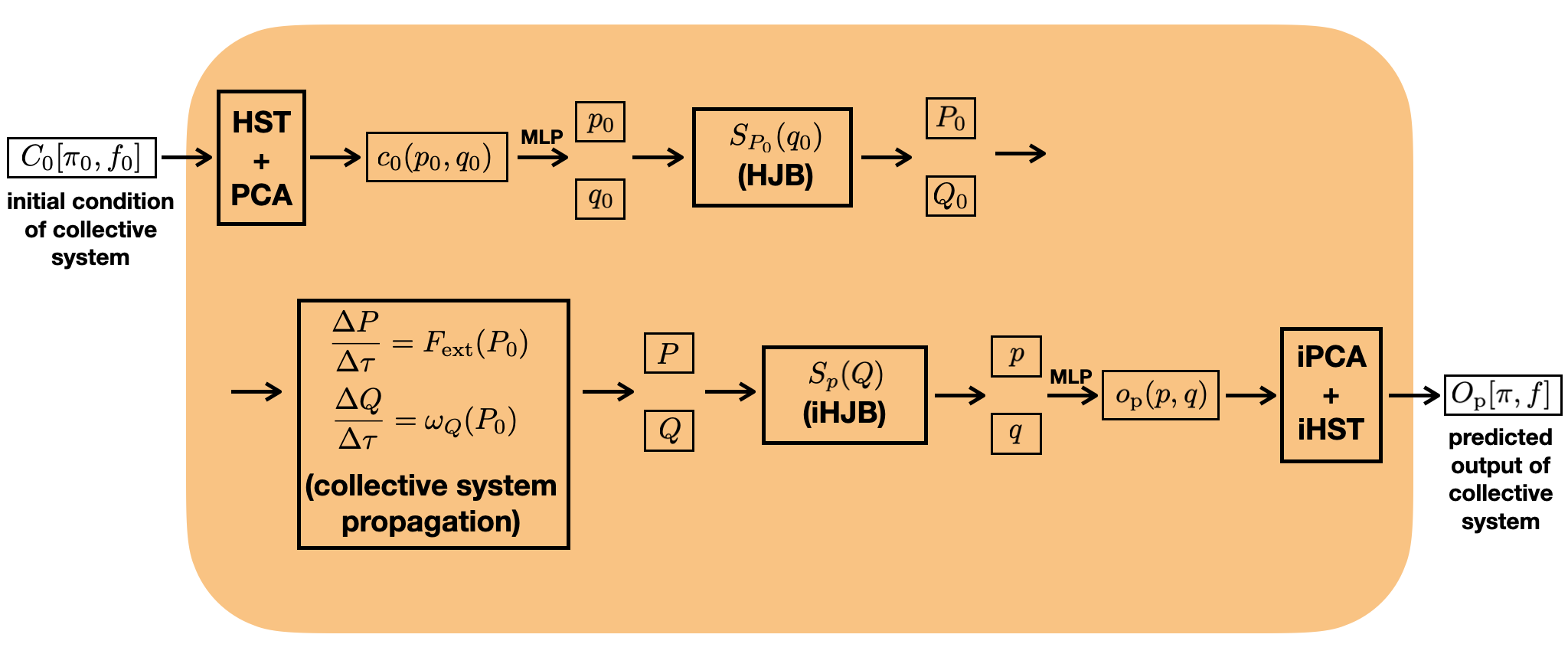}
\caption{\label{simulation.fig} Shown is the computational pipeline that simulates a collective system given the initial conditions.  The input is the initial condition of the collective system $C_0[\pi_0(x),f_0(x)]$.  The output is the predicted output the system $O_p[\pi(x),f(x)]$.}
\end{figure}

Given this theory, we move on to the practical applications of simulation and control of collective systems.  This application will use the concepts of Artificial Intelligence,\citep{hastie09,sugiyama15,goodfellow16} as interpreted by Glinsky,\citep{glinsky.24b,glinsky.24d} to approximate both the generating functional, $S_p[f(x)]$, and the generating function, $S_P(q)$.  The computational pipeline to characterize and simulate the collective system is shown in Fig.~\ref{simulation.fig}.  This pipeline contains two core components and their inverses -- one to approximate the generating functional (that is, HST+PCA and iPCA+iHST), and one to approximate the generating function (that is, HJB and iHJB).  There are also some practical components in the pipeline to approximate coordinate transformation functions.

This computational pipeline predicts the output of the collective system given the initial condition.  It does this by:
\begin{enumerate}
    \item setting the initial condition functional of the field and the co-field functions, $C_0[\pi_0(x),f_0(x)]$,
    \item transforming the initial condition, using the Heisenberg Scattering Transformation (HST) and a Principal Component Analysis projection (PCA),  from a functional of the field and co-field functions to a function of the basic state and co-state variables, $C_0[\pi_0(x),f_0(x)] \to c_0(p_0,q_0)$,
    \item  transforming from the initial condition function to the basic state and co-state variables, using a Multi-Layer Perceptron (MLP) with ReLU activation, $c_0(p_0,q_0) \to (p_0,q_0)$ ,
    \item  transforming from the basic state and co-state variables to the fundamental state and co-state variables, using the transformation generated by the solution of the Hamilton-Jacobi-Bellman equation (HJB), $(p_0,q_0) \to (P_0,Q_0)$,
    \item propagating the fundamental state and co-state variables using Eqns.~\eqref{finite.P.eqn} and \eqref{finite.Q.eqn}, $(P_0,Q_0) \to (P,Q)$,
    \item transforming from the fundamental variables to the basic variables using the iHJB transformation, $(P,Q) \to (p,q)$,
    \item transforming from the basic variables to the output function of the basic variables using a MLP with ReLU, $(p,q) \to o_\text{p}(p,q)$,
    \item transforming from the output function of basic variables to the output functional of the field and co-field functions using the iPCA and the iHST, $o_\text{p}(p,q) \to O_\text{p}[\pi(x),f(x)]$,
    \item finally, displaying the predicted output functional of the collective system, $O_\text{p}[\pi(x),f(x)]$.
\end{enumerate}
This pipeline's input is the initial condition of the collective system $C_0(x)$, and the output is the predicted output of the system $O_\text{p}(x)$.

Given this computational pipeline and a dataset, parameters of the approximation must be fit.  For the functional HST approximator, the parameters are the principle components $\left| \beta_i(z) \right>$ estimated by the PCA.  For the HJB function approximator and the coordinate transformation function approximators, the parameters are those of MLPs estimated by the optimization techniques of AI training.

The dataset is constructed by either doing an ensemble of computer simulations of the collective system or by observing the collective system.  It is helpful to apply an external force $F_{\text{ext}}$ to the system being simulated or observed to sample phase space more efficiently.  A good choice would be a dissipation or a random diffusion which samples phase space well, as the system gradually relaxes to the stable equilibriums.  It is also good to apply an external force that is constructed to keep the dynamic trajectory in the vicinity of the unstable local maximums, that is stabilizes the unstable equilibriums.  The variables that should be recorded to form the dataset are:  the initial conditions $C_0(x)$, the times $\tau$, and the outputs $O_\text{p}(x,\tau)$.

\begin{figure}
\noindent\includegraphics[width=\columnwidth]{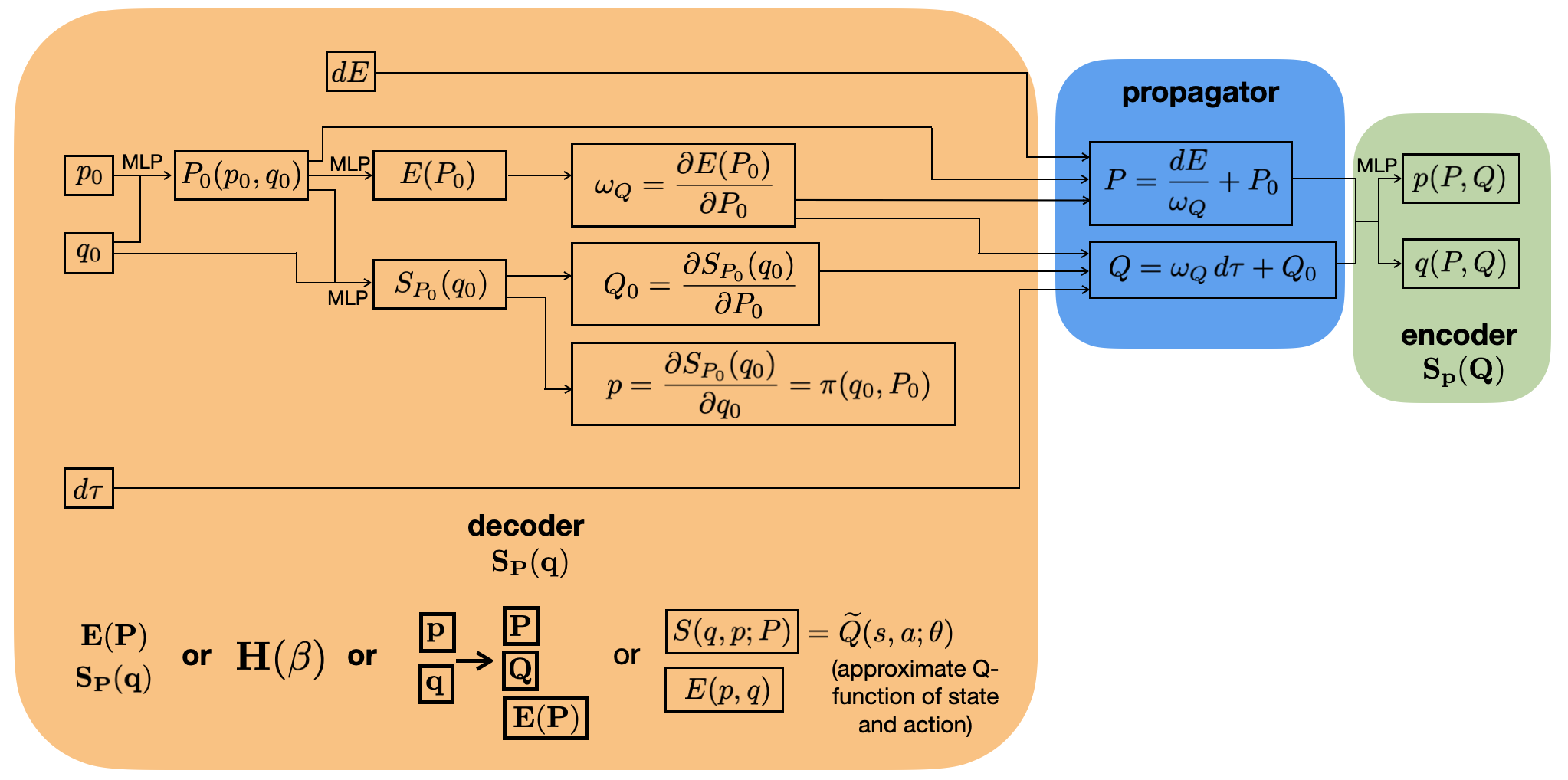}
\caption{\label{hjb.fig} The neural network architecture to estimate the solution of the Hamilton-Jacobi-Bellman equation (without resistivity).  It starts with a decoder from $(p,q)$ to $(P,Q)$ generated by $S_P(q)$ related to the imaginary part of the analytic function $H(\beta)$, including a mapping to the real part of the analytic function $H(\beta)$ given by $E(P)$, the frequency $\omega_Q(P)$, and the policy $\pi(q,P)$.  This is followed by the analytic mapping to advance from $(P_0,Q_0)$ to $(P,Q)$, then a encoder from $(P,Q)$ to $(p,q)$ generated by $S_p(Q)$.  The universal function approximator is a Multi Layer Perceptron (MLP) with ReLU.}
\end{figure}

The structure of the HST functional approximator was discussed in detail in Sec.~\ref{theory.HST.sec}, and shown in Fig.~\ref{hst.fig}.  The structure of the HJB function approximator (without the resistivity introduced by Bellman) is shown in Fig.~\ref{hjb.fig}.  It has as inputs $(p_0,q_0)$ and $(dE,d\tau)$.  The fundamental output is $(P,Q)$ then $(p,q)$, but the approximator is structured so that there are nodes corresponding to $(P_0,Q_0)$, $S_P(q)$ or the action, $E(P)$ or the energy, $\omega_Q(P)$ or the frequency, and $\pi(q,P)$ or the ``approximate policy'' $\pi_\theta(s)$ of DRL.  This enforces the structure of canonical motion generated by the $F_2$ generating function, $S(P,E;q,\tau)$, that has the form given in Eq.~\eqref{action.eqn} -- the solution to the Hamilton-Jacobi equation.  The analytic, both infinitesimal and finite, advance of $P$ and $Q$ are given by
\begin{equation}
\label{dp.eqn}
    dP= \frac{dE}{\omega_Q}  = \omega_P \, dE = F_\text{ext} \, d\tau
\end{equation}
or
\begin{equation}
\label{finite.P.eqn}
\begin{split}
    \Delta P &= P - P_0 \\
    &= \int_0^\tau{\omega_P \frac{\partial H}{\partial \tau} \, d\tau} = \int_0^{\Delta E}{\frac{dE}{\omega_Q}} \\
    &= \int_0^\tau{F_\text{ext} \, d\tau},
\end{split}
\end{equation}
and
\begin{equation}
\label{dq.eqn}
    dQ = \omega_Q \, d\tau
\end{equation}
or
\begin{equation}
\label{finite.Q.eqn}
    \Delta Q = Q - Q_0 = \int_0^\tau{\omega_Q \, d\tau},
\end{equation}
where $\omega_P \equiv 1/\omega_Q$ and
\begin{equation}
    \omega_Q(P) \equiv \frac{\partial E(P)}{\partial P}.
\end{equation}

Multi Layer Perceptrons (MLPs) \citep{goodfellow16} are used to approximate the functions.  The derivative functions are calculated by back propagating the MLPs.  It is important that Rectified Linear Units (ReLUs) are used as activation functions in the MLPs because the MLPs are approximating analytic functions which are maximally flat, but do have a limited number of singularities where the derivative is discontinuous.  MLPs with ReLUs are very good at doing this since they are universal piecewise linear approximators with discontinuities in the derivative.

Given the ensemble of inputs ($p_0$, $q_0$, $d\tau$) and outputs ($p$, $q$), minimize the ensemble average of the action $S_P(q)$, and ensemble average of the $L_2$ norm of the difference between the predicted $(p_p,q_p)$ and the simulated $(p,q)$, as well as the difference between their predicted $(\dot{p}_p,\dot{q}_p)$ and their simulated $(\dot{p},\dot{q})$ incremental derivative or vector.  This gives two scalar objectives to optimize to determine the two real-valued scalar functions $E(P)$ and $S_P(q)$, or equivalently the one complex-valued analytic function $H(\beta=P+\text{i} \, q)=E(P)+ \text{i} \, Q_P(q)$, where $Q=\partial S/\partial P$.

\begin{figure}
\noindent\includegraphics[width=\columnwidth]{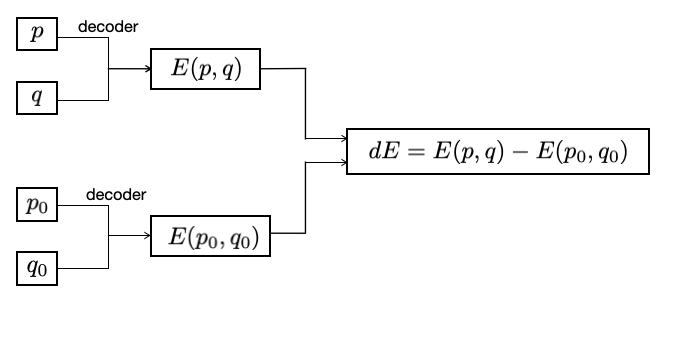}
\caption{\label{training.dE.fig} The addition to the neural network architecture to estimate $dE$ in the solution of the Hamilton-Jacobi-Bellman equation.}
\end{figure}

\begin{figure}
\noindent\includegraphics[width=\columnwidth]{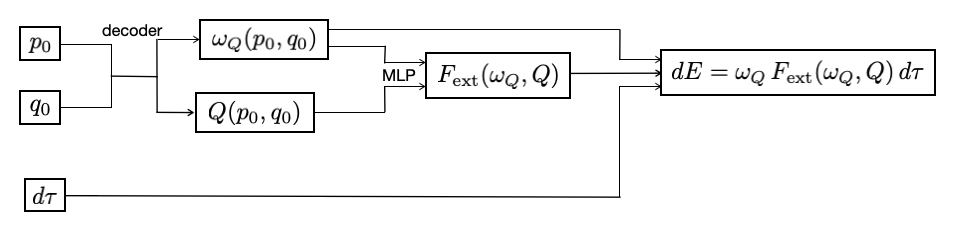}
\caption{\label{dissipation.fig} The addition to the neural network architecture to estimate $F_\text{ext}(\omega_Q,Q)$ in the solution of the Hamilton-Jacobi-Bellman equation.  The universal function approximator is a Multi Layer Perceptron (MLP).}
\end{figure}

There is a non-trivial detail in the training.  Although one has the inputs ($p_0$, $q_0$, $d\tau$) and outputs ($p$, $q$), what is $dE$?  For a conservative system with no external force being applied $dE=0$, but that is not the case with this dataset.  The solution is to use the encoder $E(p,q)$ to estimate $dE=E(p,q)-E(p_0,q_0)$, using the target outputs as an input to estimate $E(p,q)$, as shown in Fig.~\ref{training.dE.fig}.  If this workflow is being used to train a surrogate where the external force is part of the dynamics that is being modeled, a model for the external force $F_\text{ext}(\omega_Q,Q)$ needs to be estimated using an MLP so that $dE= \omega_Q \, F_\text{ext}(\omega_Q,Q) \, d\tau$, as shown in Fig.~\ref{dissipation.fig}.  If the force is resistive, diffusive friction $F_\text{ext}=-\epsilon_P \, \omega_Q$, where $\epsilon_P \ll J_0 \equiv E_0/\omega_0$.  In this case, Rayleigh's Dissipation Function can be defined
\begin{equation}
    \mathscr{F} \equiv \frac{\epsilon_P}{2} \, \omega_Q^2
\end{equation}
so that
\begin{equation}
    \frac{dE}{d\tau} = -\epsilon_P \, \omega_Q^2 = -2 \mathscr{F}.
\end{equation}
You could view the estimation of $F_\text{ext}(\omega_Q,Q)$ as an estimation of Rayleigh's Dissipation Function where
\begin{equation}
\label{thermal.force.eqn}
    \mathscr{F}(\omega_Q,Q) = - \frac{\omega_Q}{2} \, F_\text{ext}(\omega_Q,Q).
\end{equation}

Since one now has a solution of the HJB equation that has estimated the analytic Hamiltonian $H(\beta)$, one can now find the equilibriums $\beta^*$ or $P^*$ where
\begin{equation}
    \frac{\partial E(P^*)}{\partial P}=0.
\end{equation}
Given the function $E(P)$ estimated in the workflow shown in Fig.~\ref{hjb.fig}, $P^*$ can be found with a high performance root finder, both stable and unstable equilibriums.  The equilibrium or optimal policy can then be estimated as $\pi^*(q)=\pi(q,P^*)$ and the equilibrium or optimal action as $S^*(q)=S_{P^*}(q)$.

It is interesting to note that at the equilibrium points, $P^*$, the external forces can not change the system's energy because $\omega_Q(P^*)=0$ and $dE/d\tau=\omega_Q(P^*) \, F_\text{ext}=0$ for all $F_\text{ext}$.  This is because the effective mass of the collective system, $m_{\text{eff}} \sim \omega_Q^{-2}$, is infinite at an equilibrium where $\omega_Q=0$. 

The simulation workflow has some, but not all, of the elements and structure of the workflow of Glinsky and Maupin.\citep{glinsky23}  The MST and WPH were not as efficient to calculate as the HST, with $N^2$ scaling compared to the $N \log N$ of the HST.  The performance, while very good, was not excellent because the form of the MST and WPH is not that of the HST.  The inversion of the WPH (that is, iWPH) required a very slow optimization, with $N^3$ scaling, compared to the $N \log N$ scaling of the iHST.  The really good news, with respect to the workflow of Glinsky and Maupin, was that their workflow did not require the optimization of the deep convolutional network over a Banach space, an NP hard computation, and had a known simple structure for the MLP autoencoder.  The workflow did include the logarithm, and had an MLP with ReLU autoencoder, which found a ROM, but had no physical interpretation of and use for the ROM.  The result was a simulation that was $10^7$ times faster, excluding the slow iWPH, than integrating the MHD field equations, and that was trained in less than a minute on a 20 CPU core Alienware gaming computer with a Nvidia GTX 3090 graphics card.

\begin{figure}
\noindent\includegraphics[width=\columnwidth]{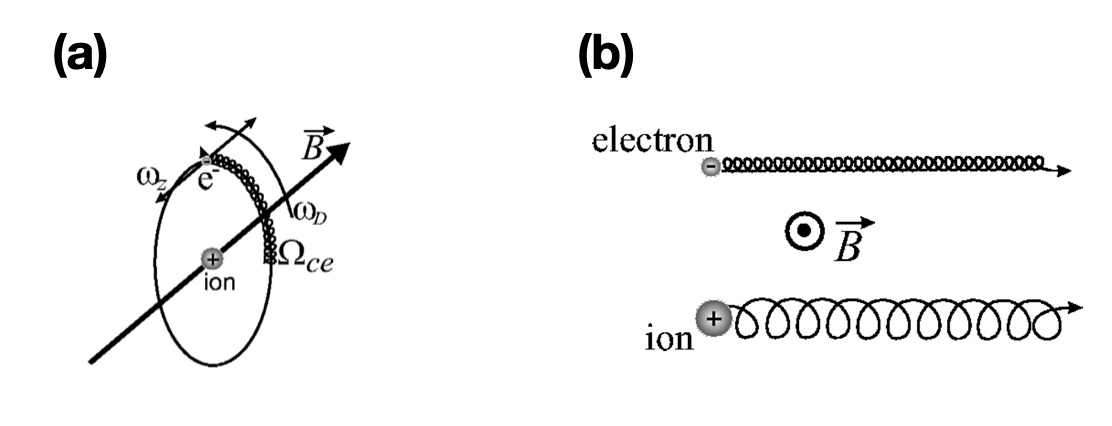}
\caption{\label{gca.dp.fig} Bound motion of: (a) the Guiding Center Atom (GCA), and (b) the Drifting Pair (DP).}
\end{figure}

It is enlightening to consider a physical system of an electron and an ion in a constant magnetic field.  It is assumed that the magnetic field is strong enough that the electron undergoes guiding center motion, that is the electron cyclotron motion is an adiabatic invariant.  It is also assumed that this magnetic field is strong enough that the motion of the electron along the magnetic field is also adiabatic.  This gives two types of bound motion.  The first is called Guiding Center Atoms (GCAs),\citep{glinsky91} where the ion is assumed to be infinitely massive.  The second is called Drifting Pairs (DPs),\citep{kuzmin04} where the electron is assumed to have no mass.  These two bound states are shown in Fig.~\ref{gca.dp.fig}.

\begin{figure}
\noindent\includegraphics[width=12pc]{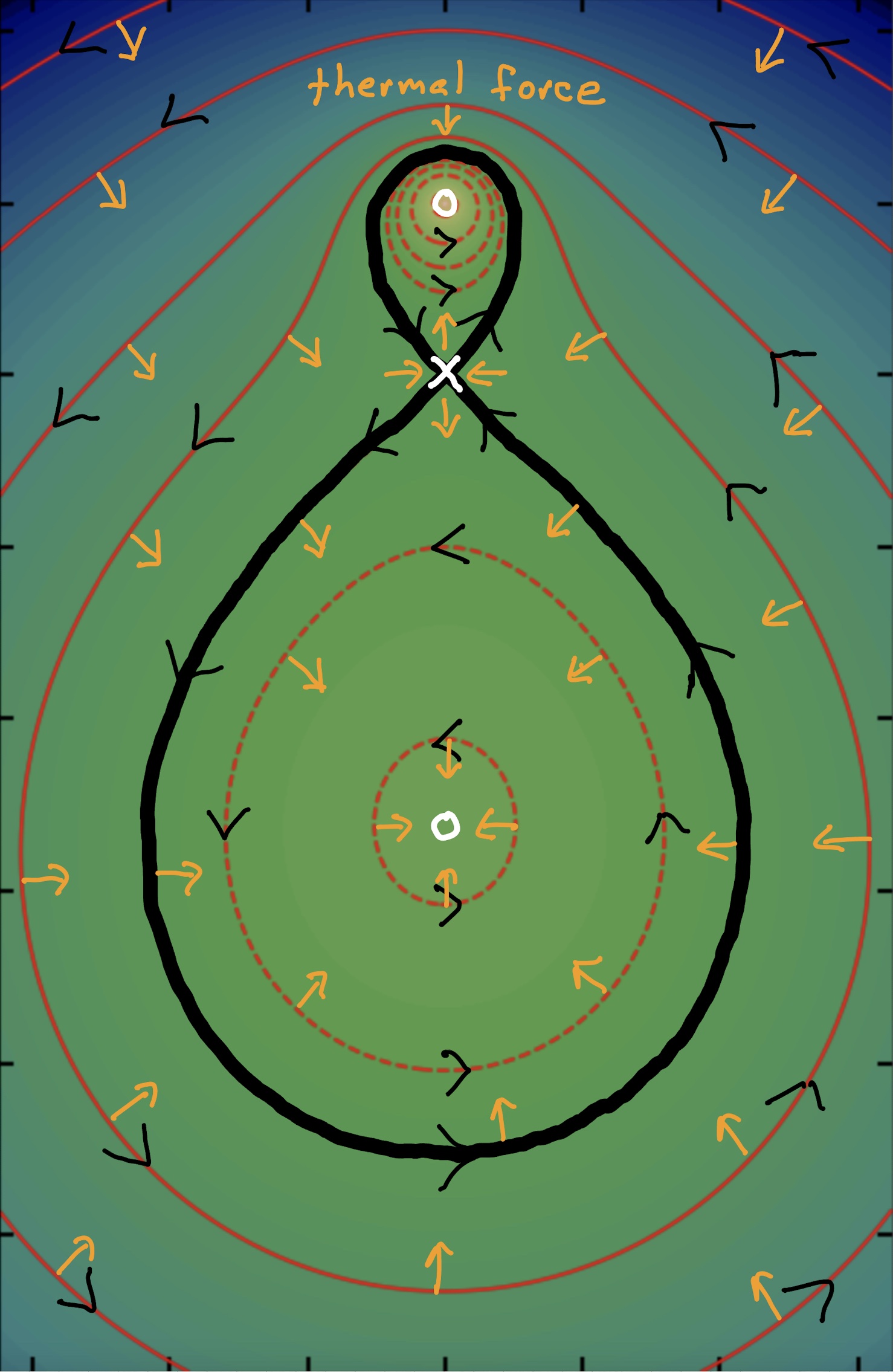}
\caption{\label{gca.dp.plot.fig} Plot of phase space trajectories shown by red lines with black arrows of an electron motion about an ion in a plane perpendicular to a strong magnetic field.  Note the boundary (thick black line) between two basins with the basin centers indicated by the white o-points, and the mountain heights with a mountain pass indicated by the white x-point.  The motion will circulate around the red lines and slowly relax in the directions shown by the orange arrows due to dissipative thermal forces.  The motion will relax from the mountain heights and eventually end up at the mountain pass (x-point), but this point is an unstable equilibrium and the motion (without control) will relax across the boundary into one of the two basins depending on how it approaches the mountain pass.  The motion will then continue to relax to the basin center at the o-point.  These are stable equilibriums.}
\end{figure}

The phase space for this system is shown in Fig.~\ref{gca.dp.plot.fig}.  This system has two stable equilibriums, shown as the o-points, and two unstable equilibriums, shown as the x-points.  One set is for the electron at $0$ and $-0.3$, and another set is for the ion at $-0.9$ and $\infty$.  There are basins for the GCA and the DP bound motions separated from the free motion by the black boundary called the separatrix.  The o-points are stable local minimums in the energy, and the x-points are unstable saddle points that are local maximums in the energy in the vertical direction.  It will take an infinite amount of time to approach the saddle points so that they will be metastable.  An external thermal force coming from a heat bath will exert a force perpendicular to the motion towards lower energy.  This will cause the motion to descend from the mountain top, spiraling around the mountain until the mountain pass, that is the saddle point, is reached.  It will take a long time to reach the mountain pass, and an equally long time to move away from the mountain pass.  The motion will then fall into one of the two basins of bound motion and spiral down to the valley center, that is the stable equilibrium.  The motion will eventually wander around the valley center in thermal equilibrium

It is desirable to have the collective system operate at the x-point since the performance, that is energy, of the x-point is better than the o-points.  The problem is that the equilibrium is unstable and will disrupt.  This sets the stage for ponderomotive stabilization of the unstable equilibriums.

The ponderomotive method of stabilization \citep{landau76} can be intuitively understood this way.  The applied ponderomotive force vibrates the collective system more when the system evolves in a undesirable way, and it vibrates the collective system less when it evolves in a desirable direction.  The collective system does not like to be vibrated, so a conservative ponderomotive potential is established that leads to a ponderomotive force away from the undesirable states and towards the desirable states.  This is because the effective mass is $\infty$ at the unstable equilibrium, and decreases as the collective system moves away from the equilibrium.  Since the force is constant, the force will move the collective system more, the further the collective system is from the equilibrium.  An alpine valley has been created at the mountain pass.

\begin{figure}
\noindent\includegraphics[width=\columnwidth]{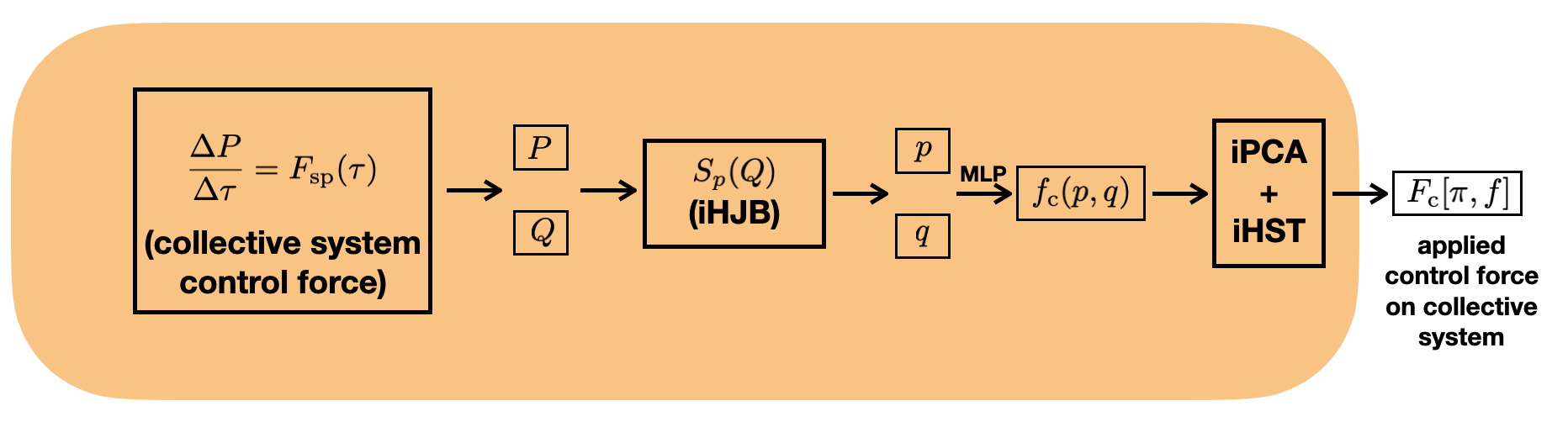}
\caption{\label{control.fig} Shown is the computational pipeline that ponderomotively controls a collective system by applying a rapidly oscillating control force given by Eq.~\eqref{stable.pondermotive.eqn}.  The output is a control force field to apply to the collective system $F_\text{c}[\pi(x),f(x)]$.}
\end{figure}

The computational pipeline that controls the collective system, both stabilizing an unstable equilibrium and cooling the collective system to the unstable equilibrium, is shown in Fig.~\ref{control.fig}. The fundamental external force that needs to be applied to control the system is given by
\begin{equation}
\label{stable.pondermotive.eqn}
    \boxed{F_\text{sp}(\tau) = f_0 \, \text{e}^{\text{i} \, \omega_\text{sp} \tau} + \omega_0 \, \varepsilon_P,}
\end{equation}
where $\omega_0 \ll \omega_\text{sp}$, $J_0 \, \omega_0 \lesssim f_0 \ll J_0 \, \omega_\text{sp}$ (so that the ponderomotive force is large but the motion is small),  $J_0 \equiv E_0/\omega_0$, and $\varepsilon_P$ is a random $\Delta P$ of size $\epsilon_P \ll J_0$ taken every $2 \pi/\omega_0$.  This $F_\text{sp}(\tau)$ force is not dependent on $P^*$ or $P$, just the time invariant mapping generated by $S_P(q)$, that is $p(P,Q)$ and $q(P,Q)$, and the functional transformation iPCA+iHST generated by $S_p[f(x)]$, that is $f[p(\tau),q(\tau)](x)$ and $\pi[p(\tau),q(\tau)](x)$.  This pipeline is vibrating the collective system with a high frequency carrier signal modulated with the characteristic ``local spectrums'' of the collective system, that is $\beta^*_i(z)$.  This is a set of measure zero compared to the Hilbert space of possible ``local spectrums''.

This computational pipeline controls the collective system.  It does this by:
\begin{enumerate}
    \item calculating the control force, $F_\text{sp}(\tau)$, that needs to be applied, $(P_0,Q_0) \to (P,Q)$,
    \item transforming from the fundamental variables to the basic variables using the iHJB transformation, $(P,Q) \to (p,q)$,
    \item transforming from the basic variables to the control function of the basic variables using a MLP with ReLU, $(p,q) \to f_\text{c}(p,q)$,
    \item transforming from the control function of basic variables to the control functional of the field and co-field functions using the iPCA and the iHST, $f_\text{c}(p,q) \to F_\text{c}[\pi(x),f(x)]$,
    \item finally, applying the control functional to the collective system, $F_\text{c}[\pi(x),f(x)]$.
\end{enumerate}
This pipeline's output is a control force field to apply to the collective system $F_\text{c}(x)$.

Modern control systems theory, pioneered by Bellman and Kalman, is for individual systems in the $(p_i,q_i)$ domain.  They have isolated singularities $\beta_i^*$ (that is, natural frequencies).  The complication with collective systems of many individual systems is that they evolve in the $[\pi_i(x),f_i(x)]$ domain with singularity spectrums, $\left| \beta_i(z) \right>$ or $\beta^*_i(z)$, (i.e., natural frequency spectrums).  What the HST+PCA is calculating are those singularity spectrums.  In order to control a collective system, you not only need to know $\beta_i^*$, but also $\beta^*_i(z)$.

More details on the mathematics and concepts presented in this section, as well as other ways of using them to control the system, can be found in \citet{glinsky23b} and \citet{glinsky.24e}.

\section{Systems-of-Systems}
\label{system.sec}
To couple a collective system to subordinate as well as superordinate collective systems, the theory must be expanded to include the coupling constants as additional fields $f_\text{c}(x)$, and co-fields $\pi_\text{c}(x)$.  Coupling constant variation will also need to be added to the AI training dataset of Sec.~\ref{sim.sec}.  The HST+PCA can be used to transform $[\pi_\text{c}(x),f_\text{c}(x)]$ to $\mathbb{C}^{n_\text{c}}$, where $n_\text{c}$ is the number of coupling constants, just as $[\pi(x),f(x)]$ were transformed to $\mathbb{C}^n$, where $n$ is the number of fields.  The complex function $C(\beta)$ can then be estimated by the HJB.  Finally, the $\mathscr{C}$ group action can be primarily quantized in the $C$ direction, and the $\text{Ad}(\mathscr{C})$ group action can be secondarily quantized in the $dC$ direction.  

\begin{figure}
\noindent\includegraphics[width=15pc]{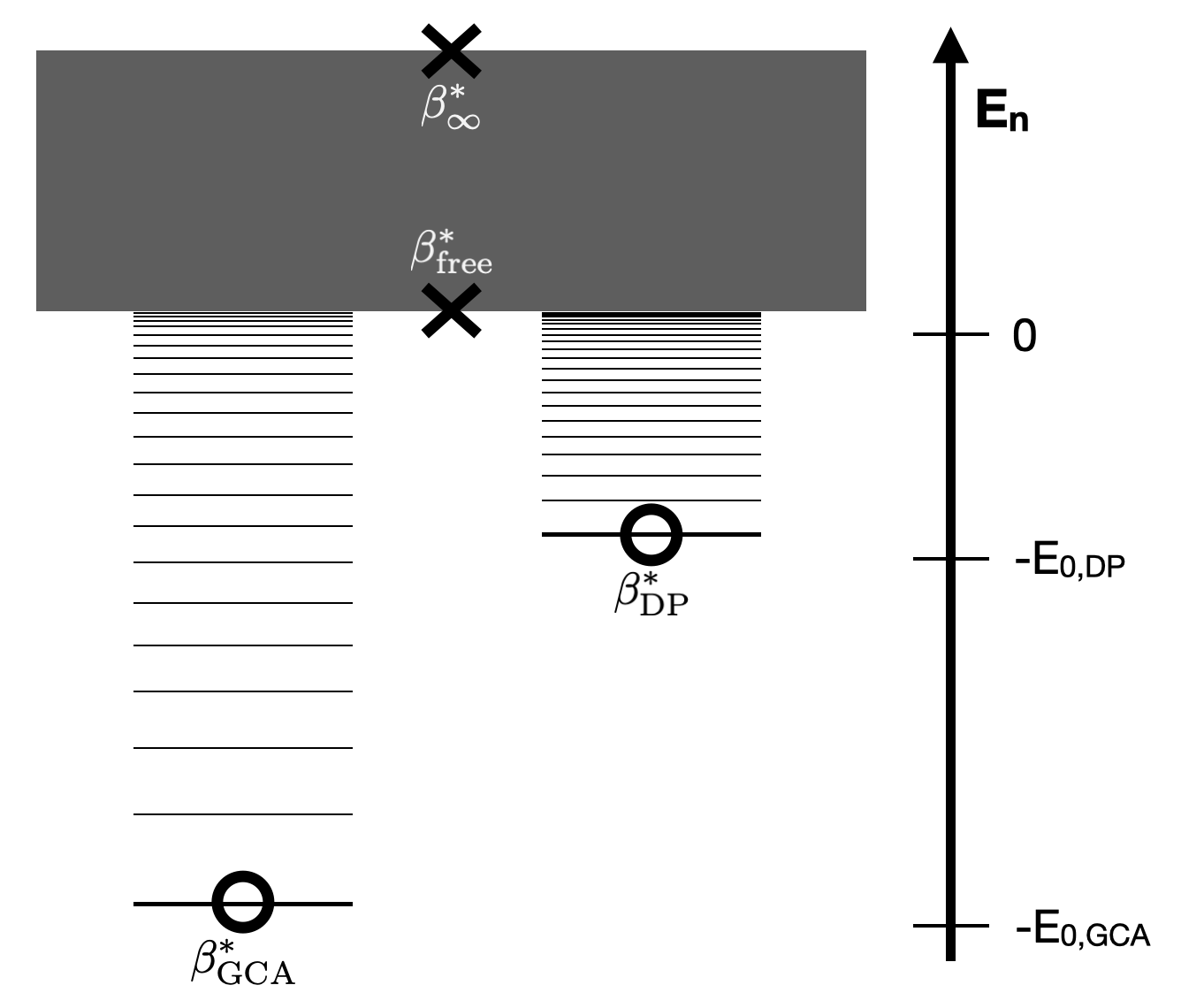}
\caption{\label{quantum.levels.fig} Quantum energy level diagram corresponding to $H(\beta)$ of Fig.~\ref{gca.dp.fig}}
\end{figure}

\begin{figure}
\noindent\includegraphics[width=15pc]{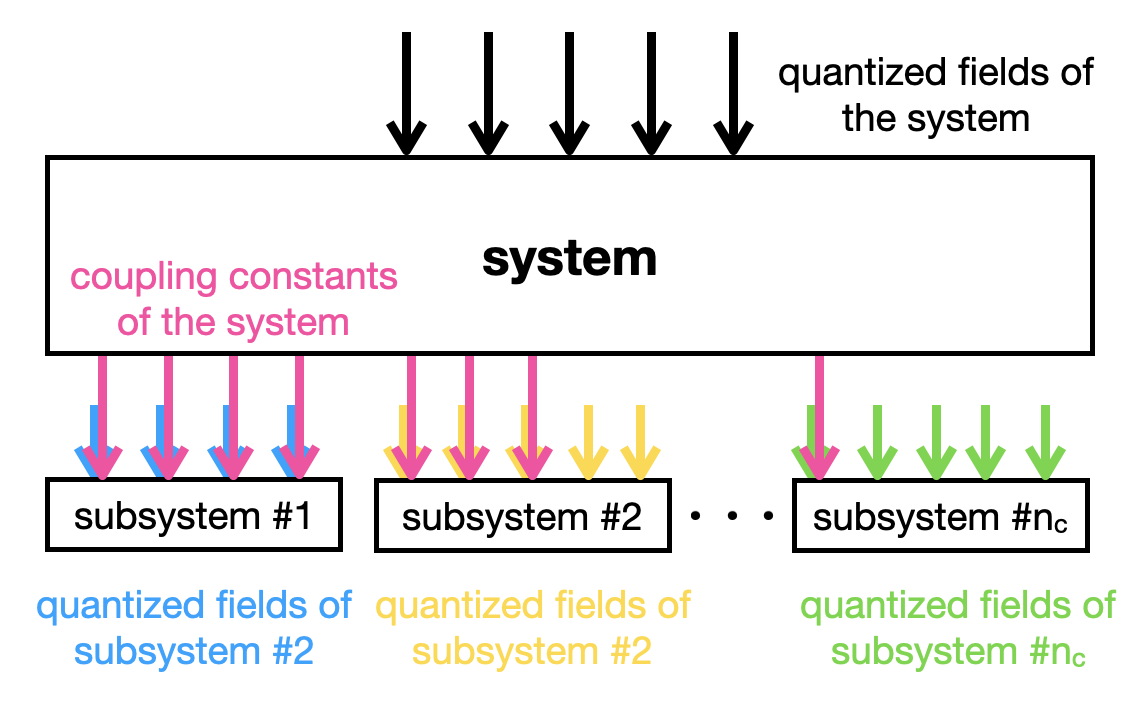}
\caption{\label{systems.fig} Coupling of the collective system to superordinate and subordinate collective systems.}
\end{figure}

The collective system is coupled to the superordinate systems via the quantized $dH$ direction, whose quantum levels and topology are determined by the first quantization of the collective system.  An example of the organization of the energy level structure, corresponding to the $H(\beta)$ of Fig.~\ref{gca.dp.plot.fig}, is shown in Fig.~\ref{quantum.levels.fig}.  Note that the energy levels for the GCA and DP are integer quantized because the motion is bounded, and the energy levels for ``free'' are rationally quantized because the motion is unbounded.  The superordinate system, if assumed to be large and in thermal equilibrium with a thermal bath with temperature $T$, will evolve the statistical distribution in the quantized $dH$ direction, with a transition kernel $K(nj,n'j')$ that satisfies detailed balance and has a thermal equilibrium of $\exp{(-\Delta E_{nj} / k_B T)}$ above each ground state, $\beta^*_j$, and $\exp{(-E_{0j} / k_B T)}$ for the relative population of each complex, $j$.  The thermal force is normal to the geodesics (that is, normal to the leaves of the foliation) defined by $\text{Re}(H(\beta))=\text{constant}$, that is in the $dH$ direction.  Note that there was a pragmatic resistive coupling of the collective system to the superordinate system using a thermal force, given in Eq.~\eqref{thermal.force.eqn}.  The theoretical approach of this section gives a more intuitive and flexible framework for external interactions that will be explained later in this section.  The collective system is coupled to the subordinate systems via the second quantized $dC$ direction, whose quantum levels and topology are determined by the first quantization of the subordinate systems.  A diagram of this arrangement is shown in Fig.~\ref{systems.fig}.  This is the basic building block of a directed graph of systems.

The system, without the subordinate and superordinate couplings, is a stationary system that is translationally invariant.  Since the $x$ coordinate is often one dimensional and identified with time by extending the phase space, this means that there is time reversibility -- there is no direction to time.  This changes when there are either subordinate or superordinate couplings (or both).  The time dependent changes to the coupling constants and fields, by definition, make the evolution of the system non-stationary.  The resulting lack of time translational invariance gives a direction to time, that is makes the dynamics irreversible.

The inclusion of a dissipative term in the dynamics, given in Sec.~\ref{sim.sec}, is related to using an extended Hamiltonian, but is nuanced, and can be understood much better using the theory of this section.  The inclusion of a dissipative term in the system of PDEs adds another ``coupling constant'' to the dynamics which can be varied and the PCA associated with that variation determined.  Coupling constant is put in quotes because this direction in the Hilbert space is in the covector plane $dH$, not the covector plane $dC$. It is foliating the $\mathbb{C}^{n}$ space of the second quantization of $dH$ for superordinate coupling, not foliating the $\mathbb{C}^{n_c}$ space of the second quantization of $dC$ for subordinate coupling.  It is not associated with a subordinate coupling being driven by this system but is the result of a superordinate system driving this system.  In this way, the dynamics is extended into one component of the $dH$ plane.  The nuance is that this extended direction will not generate another covector component for superordinate coupling, it is the use of an existing covector component.

\begin{figure}
\noindent\includegraphics[width=15pc]{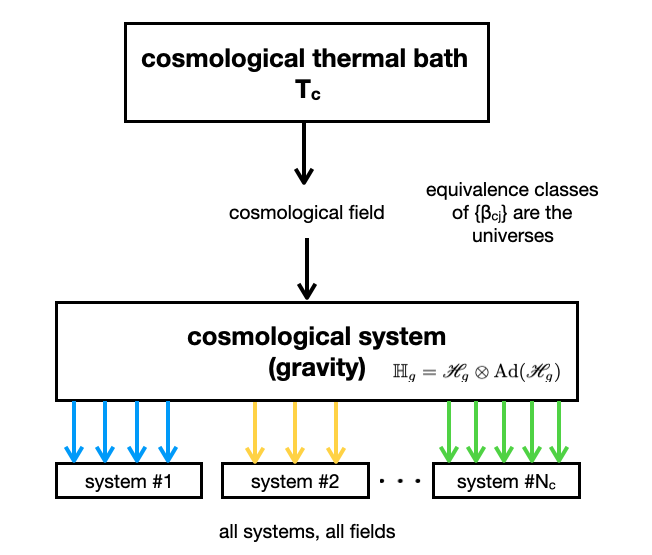}
\caption{\label{gravity.fig} Cosmological collective system couples all fields from all collective systems.}
\end{figure}

Gravity can now be introduced as the cosmological system of all systems.  All fields from all systems are the coupling constants of this cosmological system.  The gravitational symmetry group is $\mathbb{H}_g = \mathscr{H}_g \otimes \text{Ad}(\mathscr{H}_g)$, based on the gravitational Hamiltonian of harmony, $H_g$, derived by Dirac.  Assume that there is a single cosmological field with ground states, $\beta^*_{\text{c} j}$, that are simple poles and zeros (x-points and o-points).  The equivalence classes of the equilibrium states can be identified as universes (that is, basins of attraction).  The unstable equilibriums or x-points which we will call the SemiAttractors, are points at which small changes can move the cosmological collective system from one universe to another.  If the system is in the universe of a SemiAttractor, that system will, via thermal forces, move to the SemiAttractor. The stable equilibriums or o-points we will call the Attractors.  If the system moves from the universe of a SemiAttractor to the universe of an Attractor, that system will, via thermal forces, move to the Attractor.  The Attractor is a long distance from the SemiAttractor, and it will take a large amount of energy to reach the SemiAttractor.  The number of Attractors will equal the number of SemiAttractors since the system is conservative, but often one of the SemiAttractors is at infinity.  The SemiAttractor is often times a point of maximal compression, that is a Big Bang.  The equivalence class of the SemiAttractors are open universes, while the equivalence class of Attractors are closed universes.  The cosmos can be regulated by coupling the cosmological collective field to a cosmological thermal bath with cosmic background radiation temperature of $T_c$.  A diagram of the cosmological collective system is shown in Fig.~\ref{gravity.fig}.  It should be noted that gravity is quantized at the cosmological scale by the quantization of the cosmological field.  The phase space shown in Fig.~\ref{gca.dp.plot.fig} has two Attractors and one SemiAttractor.  There are three universes -- the universes of the GCAs and DPs, and the universe of free particles.

\section{Implementation of HST}
\label{code_hst.sec}
The HST is implemented in Python using pyTorch on the GPU.  It is a fast forward and inverse with $N \log N$ scaling.  This is possible because orthogonal coherent states are used as described by \citet{ali.99}.  It is implemented in the time domain with a Mother Wavelet of constant length in samples.  The Mother Wavelet is convolved with the signal.  The signal is then binated.  The convolution with the Mother Wavelet acts as an anti-alias filter for the bination.  The same wavelet is then convolved with the binated signal.  This has effectively doubled the wavelength of the filter without increasing the number of samples, leading to the $N \log N$ scaling.  The Father wavelet is matched to the Mother Wavelet and is a Gaussian-like windowing function.  The size of the window is matched to the effective scale of the convolution to have maximum spacial resolution.  Special care is taken that the set of Father Wavelets form a partition-of-unity, to preserve invertability.  Due to the Wick ordering of Eq.~\eqref{wick.eqn}, the dimensionality is preserved, but special care must be taken in posting the transform in both inverse scale, $k$, and position, $x$.  The transform is interleaved as a function of order, $m$.  Each order adds resolution.  Because of the super convergence caused by the analytic smoothness of the function that is being calculated, stopping at second order is usually sufficient.  The HST is a function of a complex variable, $k+\text{i} \, x$.

\section{Discussion and Conclusions}
\label{conclusions.sec}
Field Theory should not be ``a question of probability'', but ``a question of geometry''.  A secondary consideration of the limits on long time, $\Delta t \gtrsim 1/\omega_0$, measurement of the field leads to quantization of the stochastic probability of the field.  This paper developed a Collective Field Theory that is based on a low dimensional geodesic motion determined by the Lie group symmetry of the individual (that is, elementary particles) of the collective field.  This theory is applicable to any collective system whether that be a plasma, a fluid, a field, an economy, a society, or a language.

The transformation of the individual to the collective is taking Poisson brackets to commutators, variables to functions, functions to functionals or operators, $(p,q)$ to $[\pi(x), f(x)]$, states to wave functions and distributions, and $S_P(q)$ to $S_p[(f(x)]$.  This is not the mythic ``ascendancy'' of Quantum Mechanics.  Passing to the ``quantum'' regime, is simply passing to the stochastic regime of a dynamical or collective system, where a force must be applied to the system to measure the statistical state of the system.  The real part of the infinitesimal group generator $H(\beta)$ has a group parameter of internal evolution and equilibrium distribution $\tau$.  The imaginary part of the infinitesimal group generator $H(\beta)$ has a group parameter of external evolution and equilibrium distribution $\text{i}E/\omega$.

At the core, for the individual of the collective, is a constrained functional optimization.  This is traditionally approached via the method of Lagrange multipliers, giving the Lagrangian approach and Lagrange's equation of motion.  The problem can also be approached via the Hamiltonian perspective by making a Legendre transformation, giving the Hamiltonian approach and Hamilton's equations of motion.  In the economics and systems control literature this approach is known Pontryagins Maximum Principal of Control Systems.  This approach is known to have very useful conservation properties, a canonical structure, and sympletic geometry.  Building upon the canonical structure, is the transformational approach.  This approach solves for the generating function of a canonical transformation or flow that is the solution to the Hamilton-Jacobi equation.  This approach does not rely on the method of characteristics so that it can be applied to systems that are not integrable, that is stochastic or collective.  Although the transformational approach is the hardest to solve analytically, it is the easiest to approximate numerically as an analytic function.  In the systems control literature this approach was developed by Bellman,\citep{kalman63} then used in DRL.\citep{bertsekas96}  We also used this approach, solving for the action $S_P(q)$ of the individual of the collective, then solving for the action $S_p[f(x)]$ of the collective of individuals.  Constrained functional optimization is the core mathematics of both physics, as described by Arnold,\citep{arnold89} and economics as described by Chiang and Wainwright.\citep{chiang.05}  Just as we have moved from the elementary particle to the collective field, one can move from the economic entity to the economic collective.\citep{glinsky23b}

What is the constraint?  It is geometry, that is symmetry.  This is why we started with the basic assumption that there is at least one Lie group symmetry of the individual $\mathscr{H}$.  There can be more, as discussed in Sec.~\ref{gen.sec}.  This symmetry leads to a conserved real quantity $H(p,q)$ which is also the infinitesimal generator of the Lie group action.   We then analytically continued the real Hamiltonian $H(p,q)$ to form the complex analytic Hamiltonian $H(\beta)$, forming the complex Lie group $\mathbb{H}= \mathscr{H} \otimes \text{Ad}(\mathscr{H})$.  It is important to note that the imaginary part of the analytic Hamiltonian is simply related to the action, that is the canonical generator of the evolution of the individual or $S_P(q)$
\begin{equation}
    \text{Im}(H(\beta)) = \frac{\partial S_P(q)}{\partial P}.
\end{equation}
Remember that
\begin{equation}
    \text{Re}(H(\beta)) = E(P) = H(p,q).
\end{equation}
By solving the Hamilton-Jacobi equation, the analytic continuation of the real conserved Hamiltonian is being found.  This quantity, $S_P(q)$, is the finite canonical generator of the Lie group action.  Since $H(\beta)$ is an analytic function, knowing the singularities  of $H(\beta)$ (that is, the homology classes or topology, $\beta^*$) is equivalent to knowing $H(\beta)$.  Given the boundary condition $\beta^*$, $H(\beta)$ can be found by solving Laplace's equation.  Knowing $H(\beta)$, is knowing the topology of the Riemann surface or manifold of $H(\beta)$.   The canonical flow generated by $S_P(q)$ is simply geodesic flow on the manifold of $H(\beta)$.

This theory can also be understood from the perspective of sympletic geometry.  Let's consider the sympletic forms:  (1) the Poincaré 1-form or Lagrangian, 
\begin{equation}
    \lambda^1 = p \, dq - H \, d\tau
\end{equation}
and $\text{generator or action}=S=\int{\lambda^1}$, (2) the Poincaré 2-form or symplectic metric,
\begin{equation}
    \omega^2=d\lambda=dp \wedge dq-dH \wedge d\tau
\end{equation}
and $\text{Noether invariant or phase space volume}=H/\omega=\int{\omega^2}$ (weakly conserved), and (3) the Chern-Simmons 3-form, 
\begin{equation}
    \kappa^3=\lambda \wedge d\lambda + (2/3) \, \lambda \wedge \lambda \wedge \lambda 
\end{equation}
and $\text{topological invariant or index}=I=\int{\kappa^3}$ (strongly conserved, even with interactions with external collective systems, and associated with the homology classes $\beta^*$, that is the topology).  The Lagrangian perspective emanates from $\delta S = \delta \int{\lambda^1} = 0$.  The Hamiltonian perspective emanates from the weak conservation of $H/\omega=\int{\omega^2}$ infinitesimally generating a canonical flow.  The transformation perspective emanates from the strong conservation of $I=\int{\kappa^3}$, specifying the topology with $\beta^*$.

The concepts of genAI \citep{glinsky.24b} are utilized to approximate the generator of the individual motion, $S_P(q)$, and the generator of the collective motion, $S_p[f(x)]$.  In Sec.~\ref{sim.sec}, a computational pipeline was designed using architectures for the $S_p[f(x)]$ (that is the HST, shown in Fig.~\ref{hst.fig}) and for the $S_P(q)$ (that is the HJB, shown in Fig.~\ref{hjb.fig}) that can be trained with much less effort than the traditional approaches of DRL and GPTs.

The traditional approaches to genAI are closely related to the computational pipeline that we designed.  Optimal control, as done in genAI with Deep Q-Learning (DQN),\citep{mnih15} also known as DRL, is based on parametric estimations of a Q-function or value function
\begin{equation}
    \widetilde{Q}(s,a;\theta) = S_{P(p,q)}(q),
\end{equation}
where $s=q$ is the state, $a=p$ are the actions to be taken or the co-state, $\theta=P$ are the parameters of the estimator, and the approximate Q-function is the action.  Meanwhile, the generator of the flow, as done in genAI with Generative Pretrained Transformers (GPTs), is based on parametric estimation of a log-likelihood or score function
\begin{equation}
    \ln(\widetilde{\rho}_\theta(x)) = S_P(q),
\end{equation}
where $x=q$ is the coordinate, $\theta=P$ are the parameters of the estimator, and the approximate log-likelihood is the action.  When it comes to estimating the probability distributions, that is treating the system statistically or quantizing the theory, our work considers a dynamic equilibrium that is uniform in $Q$, and is an arbitrary function of $P$, such that the probability is
\begin{equation*}
    \rho(P) \, \text{d}P \, \text{d}Q,
\end{equation*}
where $\rho(P)$ is an arbitrary function.  In contrast, the methods of genAI consider a statistical equilibrium that is not only uniform in $Q$, but also a thermal distribution in $P$, such that the probability is
\begin{equation*}
    \text{e}^{-E(P)/k_\text{B}T} \, \text{d}P \, \text{d}Q.
\end{equation*}
This is true of all of the current methods of genAI, whether that be DQNs, GPTs, or Boltzmann Machines as estimated by Hopfield Networks, also known as MLPs.\citep{hopfield.82,hinton.85}

The connection of the HST detailed in Eq.~\eqref{hst.eqn} to Convolutional Neural Networks (CNNs) can be seen by identifying the iterative deep convolutional structure of the product, the nonlinear activation function $\text{i} \, \ln R_0$, and the pooling operation $\phi_{px} \star$, as shown in Fig.~\ref{hst.fig}.  It also should be noted that the real part of $\ln(z)$ is $\ln|z|$, and $|z|$ is a two sided ReLU.  The HST followed by the PCA projection to $\beta$ also can be interpreted as the Wigner-Weyl transformation done right on a manifold.\citep{case.08,wigner.32,weyl.50}

Limitations on the measurement of the system (that is, the Born Rule and the Heisenberg Uncertainty Principle) led to quantization of the stochastic probabilities of the collective field.  Averaging over $Q$ and enforcement of the periodic boundary conditions are the origin of quantization of these collective systems.  The quantization of the $\mathscr{H}$ group action is the primary (internally facing) quantization of the collective system, and yields the fermion particles (e.g., electrons and positrons for EM) of the collective system.  Secondary quantization is (externally facing) quantization of the $\text{Ad}(\mathscr{H})$ group action, yielding the field bosons (e.g., photons for EM) of the collective system.  This is the superordinate coupling of the collective system to other parent collective systems, like a heat bath.  For the subordinate coupling of the collective system, a group symmetry associated with the coupling constants was defined, $\mathscr{C}$.  Quantizing the $\mathscr{C}$ group action leads to the primary (internal) quantization of $C$, and quantizing the $\text{Ad}(\mathscr{C})$ group action leads to the secondary (external) quantization of $dC$.  This is the subordinate coupling of the collective system to other child collective systems.

Gravity is simply the superordinate collective field of all collective systems with the group symmetry $\mathscr{H}_g$.  This could be coupled to the uber cosmological heat bath.  The cosmological implications of this theory is the subject of ongoing research.

Although the motion of the collective in the $\mathbb{C}^n$ hyperplane will not be constant, it will be bounded with a variance of order $(2\pi)^n$, where $n$ is the number of fields.  When compared to the complete Hilbert space, this hyperplane will be of fractal dimension $n/N$, where $N$ is the number of $x$-values on which the field is defined.  This leads to the motion forming a cluster of precision $n/N$.  Since the collective motion is determined by the analytic function $H(\beta)$, it is equivalently specified by the topology of the complex manifold.  This topology is specified by the equivalence or homology classes of the topology, the ground states of the system $\beta^*$.  These ground states are the emergent behaviors of the system if there are energy dissipating terms in the dynamics.  If there are topological obstructions (that is topological invariants such as vorticity or helicity), the action will be prevented from going to zero and will be of dimension $n_t$, where $n_t$ is the dimension of the ground states.  An ensemble of relaxed motions will have a variance $(2\pi)^{n_t}$, and will be a manifold of fractal dimension $n_t/n$ with respect to the dynamical manifold.  The cluster of the ground states will have a precision of $n_t/N$ when compared to the Hilbert space and $n_t/n$ when compared to the dynamic manifold.  

\begin{figure}
\noindent\includegraphics[width=12pc]{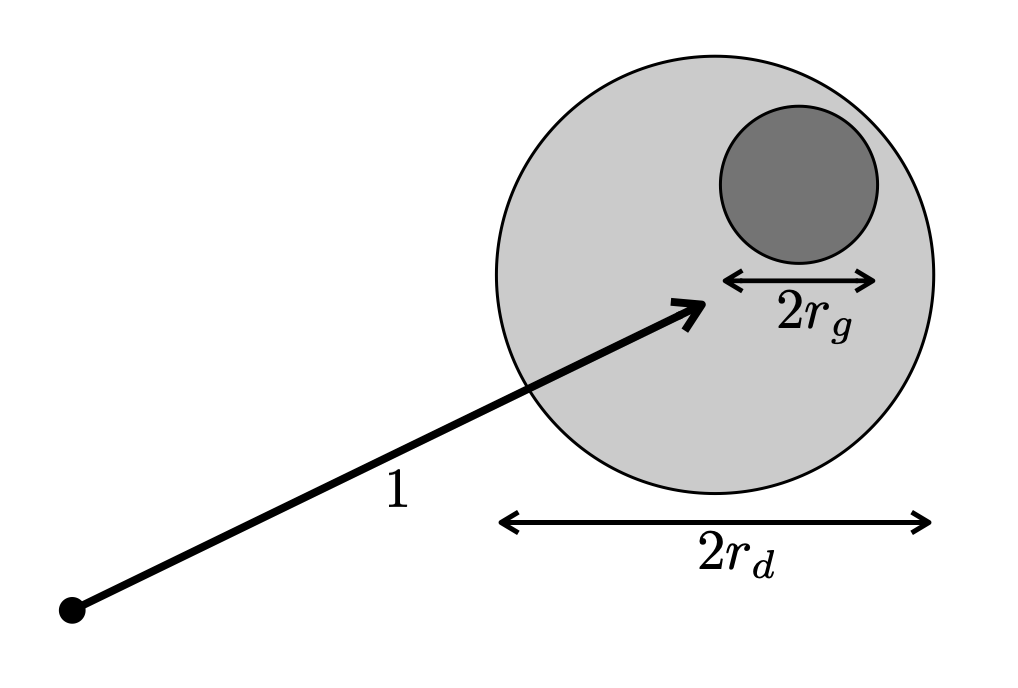}
\caption{\label{clusters.fig} Display of the precision of the cluster in $\mathbb{C}^N$ of the dynamical submanifold ($r^2_d$) and of the emergent behavior ($r^2_g$).}
\end{figure}

This is shown in Fig.~\ref{clusters.fig} for the case of $n=6$, $n_t=1$, and $N=256^2$ (that is, 2D resistive MHD with vorticity as a topological invariant.\citep{glinsky23})  The dynamical manifold will have a precision of $r_d^2 \approx 0.0001$, and the ground states will have a precision of $r_g^2 \approx 0.00002$ when compared the the Hilbert space.  The ground states will have a precision $r_g^2/r_d^2 \approx 0.2$ when compared to the dynamical manifold.

The fractal dimension of the clusters after the HST of the collective fields indicates both the number of fields, $n$, and the dimension of the ground states, $n_t$.  The fractal dimension also can be used as thresholds in algorithms as was done by Guo \emph{et al.}\citep{guo.23}  Fundamentally, the HST is a metric of collective systems.  It can be used to find the number of fields, the dimension of the ground states, and how similar two collective systems are to each other.  The use of the HST as a metric of collective systems to quantify, classify, and compare simulation to experiment, for MagLIF imploded plasma stagnation profiles, is described in Glinsky \emph{et al.}\citep{glinsky.20}

With respect to the relevance and applicability of conservative systems, two things should be noted.  Application of an external force, like a dissipative frictional force, or a diffusive Fokker-Planck thermal force or a general Boltzmann collisional force, to a conservative system does not make the system non-conservative or even change the system.  It only changes the energy of the system, that is $\Delta E = E(P')-E(P)$, where $P'=P+\Delta P$ and $\Delta P$ is the change in action.  Application of a conservative force, like a potential barrier, to a conservative system does not make the system non-conservative.  It also only changes the energy of the system, that is $\Delta E = E'(P)-E(P)$, where $E'=E+V$ and $V$ is the added potential energy.

\begin{figure}
\noindent\includegraphics[width=\columnwidth]{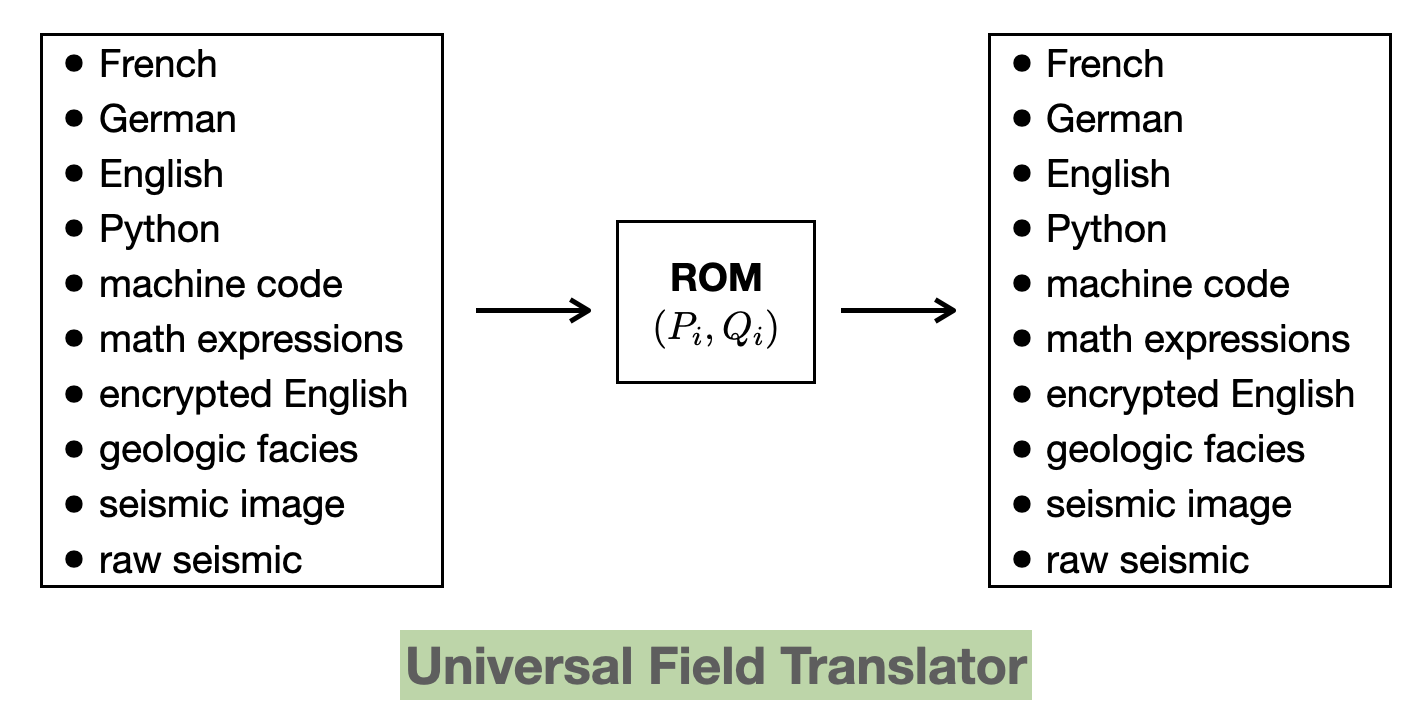}
\caption{\label{uft_1.fig} The input and output fields of the Universal Field Translator (UFT).  The UFT translates all the field types to/from a universal and minimal ROM.}
\end{figure}

\begin{figure}
\noindent\includegraphics[width=\columnwidth]{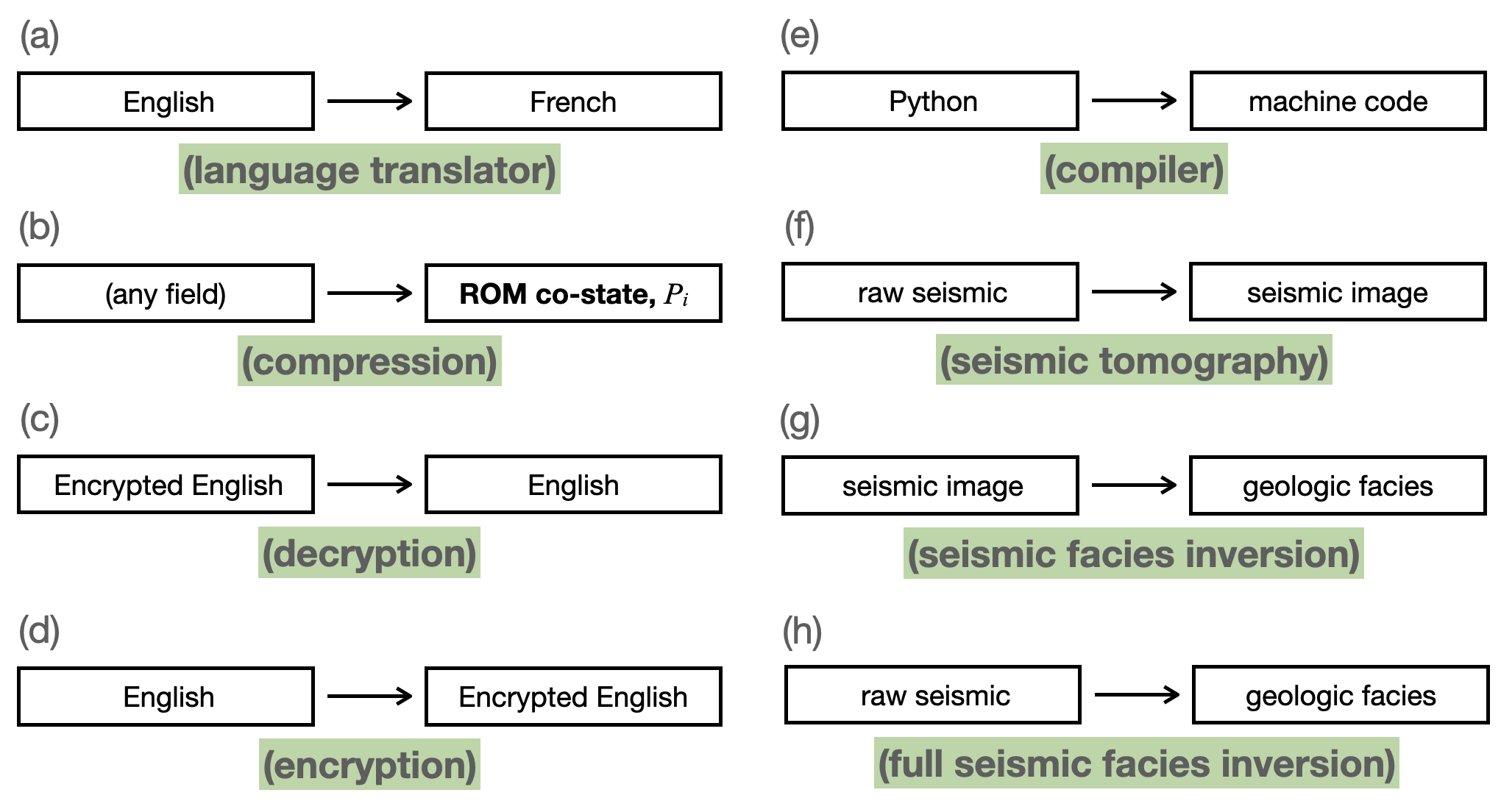}
\caption{\label{uft_3.fig} Example applications of the Universal Field Translator: (a) language translator, (b) compression, (c) decryption, (d) encryption, (e) compiler, (f) seismic tomography, (g) seismic facies inversion, and (h) full seismic facies inversion.}
\end{figure}

There are many significant applications of this theory.  As described in Sec.~\ref{sim.sec}, it can be used to simulate, efficiently with high fidelity, the evolution of the collective fields.  An ensemble of simulations can be used for Bayesian experimental and system design, as well as Bayesian data analysis (that is, data assimilation).  Ponderomotive stabilization can be used to keep the optimum system designs from disrupting.  There is a large family of applications, based on the simulation of the collective field evolution.  They are based on the fact that the fields could be:  (1) French, (2) German, (3) English, (4) Python, (5) machine code, (6) math expressions,  (7) encrypted English, (8) a geologic facies image, (9) a seismic image, or (10) raw seismic data.  The form of the genAI simulator will take one of these fields, translate it into a universal and minimal ROM $(P_i,Q_i)$, then translate it to any of the field types.  It is a Universal Field Translator (UFT) as shown in Fig.~\ref{uft_1.fig}.  Depending on what the input field and output field are, as shown in Fig.~\ref{uft_3.fig}, the UFT could be: (a) a language translator, (b) an optimal field compression, (c) a decryption, (d) an encryption, (e) a compiler, (f) a seismic tomography, (g) a seismic facies inversion, and (h) a full seismic facies inversion.

This theory unifies the four forces of nature -- the strong, the weak, the EM, and the gravitational forces.  They are all made a question of geometry, as Einstein made gravity in his General Theory of Relativity,\citep{einstein.15} where mass modifies the geometry, that is the topology, of space-time and the individual masses undergo geodesic motion given the geometry.  The complex Lie group symmetries that determine the topology are:  SU(3) for the strong force, SU(2) for the weak force, SU(1) for the EM force, and $\mathbb{H}_g$ for the gravitational force.  A Field Theory can be constructed given any Hamilton as was done for the anharmonic phononic field with $\mathbb{H}_p$, and the weather with $\mathbb{H}_w$.

The significant mathematical development of this paper is constructing a mathematically logical process of renormalization, with a physical interpretation.  This mathematically logical process is the HST which calculates the canonical generating functional $S_p[f(x)]$.  When a PCA is taken of the functional transformations, which can be viewed as localized Fourier transforms, the results are the localized spectrums that are the solutions to the RGEs.  They are the characteristic localized singularity spectrums of the Field Theory.  They also can be viewed as the S-matrix, the Mayer Cluster Expansion, the $m$-body scattering cross sections, and the $m$-body Green's functions.  The HST functional transformation is the Wigner-Weyl transformation.

This paper has been about how to discover the geometry, that is the topology, of collective fields -- \textbf{``a question of topological discovery''}.

\begin{acknowledgments}
Thanks is given to CSIRO for supporting the early parts of this research through their Science Leaders Program, and the Institute des Hautes Etudes Scientifique (IHES) for hosting a stay where the initial details of this theory were formalized and the first draft of this manuscript was written.  Thanks is also given to both St{\'e}phane Mallat and Joan Bruna for many useful discussions, and communication of many of their mathematical results before they were presented or published.  The mentorship of both Ted Frankel and Michael Freedman in topology and the geometry of physics is the foundation on which this work was constructed.  Finally, thanks is given to the University of Western Australia, John Hedditch, and Ian MacArthur for their help in understanding many of the finer points of Field Theory.

\end{acknowledgments}

\bibliography{hst_bibliography}

\end{document}